\documentclass[useAMS,usenatbib]{mn2e}
\usepackage{aas_macros}
\usepackage[dvips]{graphicx}
\usepackage{amssymb}
\title[The formation of the brightest cluster galaxies]{The formation of the brightest cluster galaxies in cosmological simulations: the case for AGN feedback }
\author[D. Martizzi et al.]{\parbox[t]{\textwidth}{Davide Martizzi$^{1}$\thanks{E-mail: martdav@physik.uzh.ch}, 
Romain Teyssier$^{1,2}$ and Ben Moore$^{1}$ }\vspace*{6pt}\\
$^{1}$Institute for Theoretical Physics, University of Zurich, CH-8057 Z\"urich, Switzerland\\
$^{2}$CEA Saclay, DSM/IRFU/SAP, B\^atiment 709, F-91191 Gif-sur-Yvette, Cedex, France}

\begin{document}

\maketitle

\label{firstpage}

\begin{abstract}
We use $500\,{\rm pc}$ resolution cosmological simulations of a Virgo--like galaxy cluster to study the properties of the brightest cluster galaxy (BCG) that forms at the center of the halo. We compared two simulations; one incorporating only supernovae feedback and a second that also includes prescriptions for black hole growth and the resulting AGN feedback from gas accretion. As previous work has shown, with supernovae feedback alone we are unable to reproduce any of the observed properties of massive cluster ellipticals. The resulting BCG is rotating quickly, has a high S{\'e}rsic index, a strong mass excess in the center and a total central density profile falling more steeply than isothermal. Furthermore, it is far too efficient at converting most of the available baryons into stars which is strongly constrained by abundance matching. With a treatment of black hole dynamics and AGN feedback the BCG properties are in good agreement with data: they rotate slowly, have a cored surface density profile, a flat or rising velocity dispersion profile and a low stellar mass fraction. The AGN provides a new mechanism to create cores in luminous elliptical galaxies; the core expands due to the combined effects of heating from dynamical friction of sinking massive black holes and AGN feedback that ejects gaseous material from the central regions.

\end{abstract}


\section{Introduction}

In the standard $\Lambda$CDM cosmological scenario, galaxy formation proceeds hierarchically and galaxy clusters are the largest and most recently assembled bound systems in the universe. The entire Hubble sequence can be found in galaxy clusters; from blue extended spirals to red massive elliptical spheroids. For this reason galaxy clusters can be considered as ideal environments for the study of galaxy formation and evolution. The brightest cluster galaxy (BCG), is usually the largest galaxy that lives at the bottom of the cluster dark matter halo. Its stars can get there in a number of ways. Some of the stars will form within the rarest peaks collapsing at high redshift - those galaxies and their dark matter halos quickly virialise through violent rapid merging and end up at the centre of the cluster. Stars can also be aquired after the cluster formed through the dynamical friction and merging with other BCGs or with the most massive satellites of the cluster. Finally, the stars may form in-situ from the cold gas that is expected to collect at the centre of the potential. The satellite galaxies themselves can undergo further evolution inside the cluster due to the interaction with the hot intracluster medium, or gravitational interactions with the cluster potential and other massive galaxies. There's growing evidence that present day massive galaxies are formed in two phase \citep{2010ApJ...725.2312O}: at redshift $z \gtrsim 2$ stars are formed within  galaxies from infalling cold gas, at redshift $z \lesssim 3$ galaxies mainly grow through accretion of stellar material. Massive early type galaxies can assemble more than half of their present day mass through dry minor mergers at $z \lesssim 2$ \citep{Naab:2009p5731}. These processes affect their star-formation histories and their morphological appearance. 

Explaining the formation of very massive BCGs at the center of galaxy clusters is one of the most challenging open problems in galaxy formation research. Standard galaxy formation models in the context of $\Lambda$CDM cosmology are affected by the so--called "overcooling problem": massive galaxies are predicted to be too bright and too blue when compared to massive galaxies in the nearby universe \citep{Borgani:2009p728}. These models find a stellar content in massive cluster galaxies that is significantly above the observed values \citep{Kravtsov:2005p702}, even if extreme supernovae feedback recipes are included \citep{Borgani:2004p1066}. One scenario that has been proposed to solve this problem involves feedback processes from Super Massive Black Holes (SMBHs), usually referred to as AGN feedback. Theoretical considerations \citep{Tabor:1993p1080, Ciotti:1997p1087, Silk:1998p941} suggest that these processes should provide enough energy to prevent gas from accumulating in the central regions of galaxy clusters thus quenching star formation. {Semi-analytical models including AGN feedback coupled with N-body simulations have shown that most of the stellar mass in present time BCGs was assembled through dry minor mergers, following a phase of quiescent star formation influenced by feedback processes \citep{DeLucia:2007p1656}. The strongest evidence supporting the existence of AGN feedback is provided by observations of X-ray cavities and radio cavities in galaxy clusters. These have been interpreted as buoyantly rising bubbles of high entropy material injected in the cluster core by jets of relativistic particles. The success of numerical hydrodynamical simulations including AGN driven buoyantly rising bubbles in reproducing masses and colours of observed BCGs has been shown by \cite{sijacki_springel06}.} Thus, AGN feedback is expected to play a significative role in shaping the properties of BCGs. 

In this paper we use two high resolution cosmological simulations of a Virgo-like cluster, with and without AGN feedback. The simulations were performed by \cite{Teyssier:2010p909} using the AMR code RAMSES \citep{Teyssier:2002p451}. \cite{Teyssier:2010p909} showed that combining high spatial resolution and the effects of AGN feedback it is possible to bring the stellar content in the cluster closer to the observed values, while a model without AGN feedback totally fails. Here, we use the same simulations to study the effect of AGN feedback on the massive BCGs that form at the center of the cluster. The main questions we want to address are: 

\begin{itemize}
 \item What are the differences between the properties of the BCG when we include AGN feedback and when we do not? We consider structural as well as
 kinematic quantities to address this problem, including ellipticities, masses, velocity dispersions, rotational velocities and stellar surface density profiles.
 \item What are the main ingredients that determine the evolution of the BCG? Massive elliptical galaxies are expected to form through a long series
 of dry minor mergers \citep{Naab:2007p3844, Naab:2009p5731}, but processes connected to SMBHs and AGN feedback may also influence their evolution.
 \item Recent results showed that it is possible to reproduce some of the properties of massive elliptical galaxies in the field and in galaxy 
 groups neglecting AGN feedback \citep{Naab:2009p5731, Feldmann:2010p1516}. Is it possible to do the same for BCGs in massive galaxy clusters? If 
 not, can the problem be solved by invoking AGN feedback?
\end{itemize}

The paper is organized as follows: the first section is dedicated to the numerical methods and the sub-grid recipes adopted for our simulations (cooling, star formation and AGN feedback), while the second section presents our results, comparing our models with observational data and with other numerical simulation studies. The final section is left for discussion.

\section{Numerical Techniques}
\label{sec:num_methods}

In this section we describe the numerical methods and the initial conditions used to model the Virgo--like cluster we will study in this paper. 
We consider two numerical cosmological simulations that have been presented by \cite{Teyssier:2010p909}. They were performed using the zoom-in technique which allows us to obtain the required effective resolution in selected regions of the computational domain. We consider a periodic cubic box of the universe of side 100 Mpc/h that has the standard $\Lambda$CDM cosmological parameters, with $\Omega_{\rm m}=0.3$, $\Omega_\Lambda=0.7$, $\Omega_{\rm b}=0.045$, $\sigma_{\rm 8}=0.77$ and $H_0=70$ km/s/Mpc. We used the \cite{Eisenstein:1998p1104} transfer function and the {\ttfamily grafic} package \citep{Bertschinger:2001p1123} in its parallel implementation {\ttfamily mpgrafic} \citep{Prunet:2008p388} to generate our initial conditions. We first ran a low resolution dark matter only simulation, then we identified dark matter halos at $z=0$. From the original halo catalog we constructed a set of candidate halos whose virial masses lie in the range $10^{14}$ to $2 \times 10^{14}$ M$_\odot$/h. Finally, we identified our final halo based on its assembly history: most of its mass is already in place at $z=1$, therefore it can be considered as relaxed at $z=0$. Its final virial mass is $M_{\rm vir} \simeq 10^{14}$~M$_\odot$, and $M_{\rm 200c}=1.04 \times 10^{14}$~M$_\odot$ or $M_{\rm 500c}=7.80 \times 10^{13}$~M$_\odot$, where indice $c$ refers to the critical density. {We stress that our aim is not to exactly reproduce the properties of the Virgo cluster, that is a relatively unrelaxed cluster, but to study the effect of AGN feedback on the formation of the central galaxy in a cluster with a similar mass. We select a relatively relaxed halo: this allows us to focus on this effect without having to deal with a very complex merger history.} The Virgo--like halo was then re-simulated at higher resolution including dark matter and baryons.

\subsection{The simulations}

The two cosmological hydrodynamical simulations considered in this paper were performed using the AMR code RAMSES \citep{Teyssier:2002p451}. The initial grid had an effective size of $2048^3$ and it has been used to extract a set of high resolution dark matter particles only in the Lagrangian volume of the halo, while a lower resolution has been used to sample the rest of the periodic box. As a result we have $22\times 10^6$ particles in the cosmological box, $19\times 10^6$ particles in the high resolution region and $8\times 10^6$ particles within the virial radius of the halo at $z=0$. This means that in the high resolution region we have a dark matter particle mass of $8.2 \times 10^6$ M$_\odot$/h and a resolution element mass of $1.4 \times 10^6$ M$_\odot$/h for the baryonic component.

The AMR grid used to solve the hydrodynamical equations was initially refined to the same level as the particle grid ($2048^3$, level $\ell=11$), but 7 more levels of refinement were considered during the run (level $\ell_{\rm max}=18$). We used a refinement criterion that allowed spatial resolution to be nearly constant in physical units; in this way the minimum cell physical size was always close to $\Delta x_{\rm min} = L/2^{\ell_{\rm max}}\simeq 500$ pc/h. The grid was dynamically refined using a quasi-Lagrangian strategy: when the dark matter or baryonic mass in a cell reaches 8 times the initial mass resolution, it is split into 8 children cells.

\begin{table}
\begin{center}
{\bfseries Mass and spatial resolution}
\begin{tabular}{|c|c|c|c|}
\hline
\hline
 $m_{\rm cdm}$&  $m_{\rm gas}$ & $m_*$ & $\Delta x_{\rm min}$ \\
  $[10^{6}$ M$_\odot$/h] & $[10^{6}$ M$_\odot$/h] & $[10^{6}$ M$_\odot$/h] & [kpc/h] \\
\hline
\hline
 $8.2$ & $1.4$ & $0.3$ & $0.38$ \\
\hline
\hline
\end{tabular}
\end{center}
\caption{Mass resolution for dark matter particles, gas cells and star particles, and spatial resolution (in physical units) for our two simulation. }
\end{table}

\subsection{Modelling the baryonic processes}

Gas dynamics is modeled using a second-order unsplit Godunov scheme \citep{Teyssier:2002p451, Teyssier:2006p413, Fromang:2006p400} based on the HLLC Riemann solver and the MinMod slope limiter \citep{Toro:1994p1151}. We assume a perfect gas equation of state (EOS) with $\gamma=5/3$. Part of the galaxy formation process has been calculated using sub-grid models, since a correct modeling of the turbulent and multiphase interstellar medium (ISM) is beyond the reach of  present-day cluster simulations. In both simulations:
\begin{itemize} 
 \item {\itshape The ISM} has been modeled using a very simple EOS for the gas 
  \begin{equation}
  T_{\rm floor} = T_* \left( \frac{n_{\rm H}}{n_*} \right) ^{\Gamma -1}
  \label{eq:eos}
  \end{equation}
  where $n_*=0.1$ H/cc is the density threshold that defines the star forming gas, $T_*=10^4$ K is a typical temperature mimicking
  both thermal and turbulent motions in the ISM and $\Gamma=5/3$ is the polytropic index controlling the stiffness of the EOS. Gas cannot cool
  below
  the temperature floor, while it can be heated above.
 \item {\itshape Gas cooling} is followed according to the \cite{sutherland_dopita93} cooling function. We take into account H, He and 
  metal cooling. Gas metallicity is advected as a passive scalar, and 
  is self-consistently accounted for in the cooling function. We also considered the effect of the standard homogeneous UV background of 
  \cite{Haardt:1996p1167}, but we modified the starting redshift, extrapolating the average intensity from $z_{\rm reion}=6$ to 
  $z_{\rm reion}=12$.
  This extrapolation is justified by the fact that early reionization is expected in proto-cluster regions \citep{Iliev:2008p1200}.
 \item {\itshape Star formation} is implemented using the following simple model. We create new star particles in cells with gas density larger than $n_*$. 
  The mass of the star particles depends on resolution; in the present case we have chosen $3\times10^5$ M$_\odot$/h. The formation rate of star particles is given by 
   \begin{equation}
   \dot \rho_{*} = \epsilon_* \frac{\rho_{\rm gas}}{t_{\rm ff}}~~~{\rm with}~~~t_{\rm ff}=\sqrt{\frac{3\pi}{32G\rho}}
   \end{equation}
  where $t_{\rm ff}$ is the local free-fall time of the gaseous component and $\epsilon_*=0.01$ is the star formation efficiency.
 \item {\itshape Supernovae feedback} is implemented in the code. A 10\% mass
  fraction of each star particle is ejected in supernovae explosions after 10 Myr.
  We assume that the supernova energy is $10^{51}$ erg. We also 
  include metal enrichment from supernovae: one M$_\odot$ of metals per 10 M$_\odot$ average progenitor mass is ejected in the ISM. {This value for the metal yield has been 
  chosen to match, on average, the metal enrichment from massive stars \citep{1995ApJS..101..181W}. Supernovae 
  feedback is modeled using the "delayed cooling" scheme \citep{Stinson:2006p1402}, i.e. we shut down gas cooling for 50 Myr in the cells surrounding a star particle going supernova. 
  Since regions where feedback acts are typically very dense, this technique prevents gas from fastly radiating away the feedback energy. See \cite{2011MNRAS.410.1391A} for 
  more details.}
\end{itemize}

It has been shown that these galaxy formation recipes are able to successfully reproduce the properties of spiral galaxies in the field  \citep{Mayer:2008p1478, Governato:2009p1455, Governato:2010p1442, 2011MNRAS.410.1391A} as well as 
other observed galaxy properties like the Kennicutt-Schmidt law, star formation rates, galactic winds \citep{Dubois:2008p393, Devriendt:2010p5266, 2011MNRAS.410.1391A}. On galaxy group scales the same recipes are less successful in reproducing the observed properties \citep{Feldmann:2010p1516}, while on galaxy cluster scales they fail \citep{Borgani:2004p1066, Kravtsov:2005p702, Borgani:2009p728}. In simulated galaxy clusters similar sets of phenomenological models produce overcooling of gas that leads to the formation of a higher fraction of stars than observed in real systems \citep{Borgani:2009p728}. {It has been argued that star formation is quenched in halos more massive than $M_{\rm c} \simeq 6 \times 10^{11}$ M$_\odot$  \citep{Cattaneo:2006p990}, and this mass threshold is thought to be related to the stabilisation of gas accretion shocks and to the transition from cold to hot gas accretion \citep{Birnboim:2003p1535}, although additional physical processes are required to prevent overcooling. The favoured theoretical scenario proposed to solve the overcooling problem involves the role of AGN feedback in halos of mass $M>M_{\rm c}$. Gas accretion favours AGN activity and the related feedback on the gas \citep{Cattaneo:2006p990}.} AGN feedback is expected to provide enough energy to heat up gas in halos and partially blow it away, thus preventing further star formation. The analysis of cosmological simulations of galaxy groups and clusters including AGN feedback like the ones of \cite{Puchwein:2008p767} support this scenario.

The two simulations of the Virgo--like cluster we are considering differ substantially. The first run has been performed using only the galaxy formation recipes described above, without considering the presence of SMBHs and neglecting AGN feedback, therefore after much thought of a clever acronymn, we call it the AGN-OFF run. In the second run we include SMBHs and we take into account AGN feedback; we call it the AGN-ON run. 

\subsection{SMBH growth and AGN feedback in the simulations}

The seeds for SMBHs formation are thought to be either Pop III stars \citep{Madau:2001p3493}, or a result direct collapse of baryonic material within low angular momentum halos \citep{Bromm:2003p3495, Begelman:2006p3499}. In both cases the seed SMBHs are expected to grow relatively quickly to $M_{\rm BH,s}=10^5$  M$_\odot$ when they will start to interact with the environment and self-regulate their gas accretion rate. The fact that this black hole mass is at least one order of magnitude lower than the minimum SMBH mass observed in the $M_{\rm BH}-\sigma$ relation \citep{Gebhardt:2000p1011, Gultekin:2009p475} allows us to consider $M_{\rm BH,s}$ as the prototypical seed SMBH mass. 

In our model, we use sink particles to simulate SMBHs, following the prescription of \cite{Krumholz:2004p1079}. When the following conditions are met, we create a new SMBH in the simulation:
\begin{itemize}
\item The stellar density has to be greater than 0.1 H/cc ($2.4\times 10^6$M$_{\odot}$ kpc$^{-3}$). This ensures that SMBHs form in stellar systems.
\item The stellar 3D velocity dispersion has to be greater than 100 km/s. With this condition we require that the line--of--sight velocity dispersion is $\sigma_{\rm 1D} \geq 60$~km/s, in agreement with the observed $M_{\rm BH}-\sigma$ relation.
\item The gas density has to be greater than 1 H/cc. With this condition we are sure that seed black holes form in the nuclear region of star forming disks.
\item No other sink particle is present within 10 kpc. No new seed SMBH will be created within 10 kpc from an old SMBH residing at the center of a galaxy. 
\end{itemize}   
Each seed SMBH has a fixed mass $M_{\rm BH}=10^5$  M$_\odot$ and a fixed radius $r_{\rm sink}=4 \Delta x \simeq 2$ kpc, where $\Delta x$ is the spatial resolution in physical units. We assume that the SMBH mass is homogeneously distributed inside a sphere of radius $r_{\rm sink}$, and we add this density distribution to the total mass density when solving the Poisson equation. Like dark matter and star particles, sink particles are advanced in time by interpolating the gravitational force back to the sink position using the inverse CIC scheme. 

{A key ingredient in our simulations is AGN feedback from SMBHs. On the theoretical side, it is thought that AGN feedback energy could be transported through the gas by several processes, e.g. radiative feedback \citep{Ciotti:2001p2231} or strong shocks \citep{Begelman:2004p2233, 2011arXiv1108.0110D, 2011MNRAS.411..349G}. Additional feedback may come from production of cosmic rays \citep{Bruggen:2002p2160, Chandran:2007p2323}. On the theoretical side, all these processes are very complex to model self-consistently and therefore are very challenging to be reproduced in hydrodynamical simulations. Additional difficulties include the very high spatial resolution required to study these processes in a detailed way, although it is not only a matter of resolution but also of proper physical modeling. Even in state-of-the-art cosmological hydrodynamical simulations it is not possible to model these processes in a self-consistent way, so that a phenomenological treatment of the problem must be chosen. }

Mass accretion onto SMBHs and AGN feedback are implemented using a modified version of the \cite{Booth:2009p501} model which was originally developed for SPH simulations. {This model is a modified version of that proposed by \cite{Springel:2005p63}.} We compute the mass accretion rate onto each SMBH using a modified Bondi-Hoyle formula
\begin{equation}
\label{bondi_formula}
\dot{M}_{\rm BH}=\alpha_{\rm boost} \frac{4\pi{\rm G}^2M_{\rm BH}^2 \rho}{ (c_{\rm s}^2+u^2)^{3/2}}
\end{equation}
where $\rho$, $c_{\rm s}$ and $u$ are the average gas density, sound speed and relative velocity within the sink radius, all computed following the approach of \cite{Krumholz:2004p1079}. The parameter $\alpha_{\rm boost}$ was introduced by \cite{Springel:2005p63} to account for unresolved multiphase turbulence in the SMBH environment, and it's value was originally chosen as constant. Numerical studies by \cite{Booth:2009p501} show that $\alpha_{\rm boost}$ should instead be considered as a function of the local density: it should be close to unity in low density regions, it should increase in high density regions, in order to match the subgrid model used for the unresolved turbulence in the disks. As proposed by \cite{Booth:2009p501}, we adopt 
\begin{eqnarray}
\nonumber
\alpha_{\rm boost}&=&\left( \frac{n_{\rm H}}{n_*} \right)^2~~~{\rm if }~~�n_{\rm H} > n_* = 0.1~{\rm H/cc,}\\
\alpha_{\rm boost}&=&1~~~{\rm otherwise.}
\end{eqnarray}
We stress that the choice for this particular form of $\alpha_{\rm boost}$ is strictly dependent on the chosen EOS for the gas. 

Equation \ref{bondi_formula} does not provide any upper limit for the accretion rate, however it cannot exceed the Eddington limit 
\begin{equation}
\label{eddington_formula}
\dot{M}_{\rm ED}=\frac{4\pi{\rm G}M_{\rm BH} m_{\rm p}}{\epsilon_{\rm r} \sigma_{\rm T}c}~~~�{\rm with}~~\epsilon_{\rm r} \simeq 0.1{\rm .}
\end{equation}
where $\epsilon_{\rm r}$ is is the efficiency at which accreting gas rest mass energy is converted into radiation. To enforce this upper limit we always set the accretion rate to
\begin{equation}
\label{accretion_formula}
\dot{M}_{\rm acc}=\min (\dot{M}_{\rm BH},  \dot{M}_{\rm ED})
\end{equation}
 At each time step, a total gas mass of $\dot{M}_{\rm acc} \Delta t$ is removed from all cells within the sink radius, with the same weighting
 scheme as the one used to define average quantities \citep{Krumholz:2004p1079}. In order to prevent the gas density from vanishing or becoming
 negative, we remove no more than 50\% of the gas at each time step.

\begin{figure}
    \includegraphics[width=0.5\textwidth]{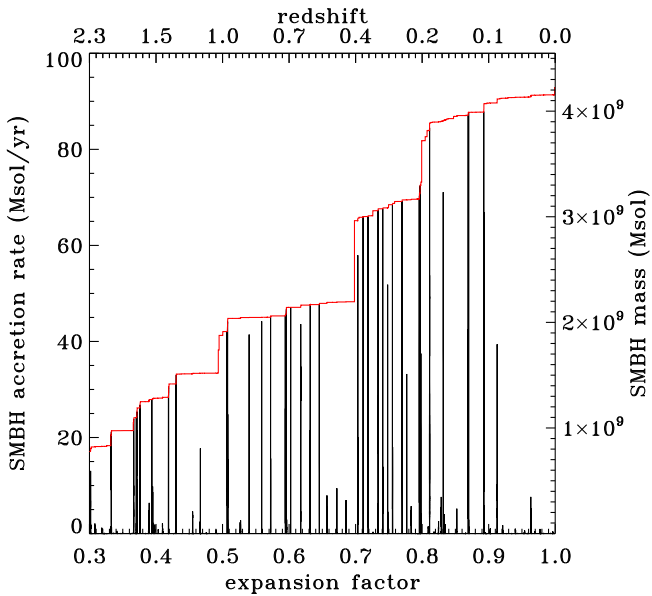}
  \caption{ Time evolution of the SMBH mass in red, and of its instantaneous accretion rate in black. }
  \label{fig:BHacc}
\end{figure}

In our simulations, AGN feedback is implemented in a way that allows self-regulated SMBH mass growth \citep{Sijacki:2007p1032, Cattaneo:2007p420, Booth:2009p501}, by injecting thermal energy directly into the gas surrounding the black hole. If the black hole mass is too low, the amount of energy injected in the gas will not be able to heat it up, so it will remain cold and will accrete onto the SMBH at rates close to Eddington limit. In this case the SMBH growth can proceed exponentially with time, with e-folding time scale equal to the Salpeter time 
\begin{equation}
\label{salpeter_time}
t_{\rm S}=\frac{\epsilon_{\rm r}\sigma_{\rm T}c}{4\pi G m_{\rm p}}\approx 45 \hbox{Myr}
\end{equation}
Feedback will become more efficient as long as the SMBH continues to grow, until the injected energy is enough to unbind the gas surrounding the black hole. In this regime accretion will become Bondi-Hoyle limited and mass growth will proceed more slowly. Figure~\ref{fig:BHacc} shows the time evolution of the accretion rate of the most massive SMBH in the simulated region. Short bursts of Eddington limited accretion are followed by long quiescent epochs of Bondi-Hoyle limited accretion, self-regulated by SMBH feedback. 

 At each time step we compute the thermal energy injected in the gas surrounding each black hole as
\begin{equation}
\Delta E = \epsilon_{\rm c} \epsilon_{\rm r}\dot{M}_{\rm acc}c^2 \Delta t {\rm .}
\label{enerdump}
\end{equation}
where $\epsilon_{\rm c}$ is the coupling efficiency, i.e. the fraction of radiated energy that is coupled with the surrounding gas. The correct value for $\epsilon_c$ can be set requiring the simulations to reproduce the observed $M_{{\rm BH}}-\sigma$ relation; we use the fiducial value $\epsilon_{\rm c} \simeq 0.15$ \citep{Booth:2009p501}. The energy $\Delta E$ is not immediately injected in the gas, but its accumulated and stored in a new variable $E_{\rm AGN}$, so that we can avoid the gas instantly radiating away this energy via atomic line cooling. We release this energy within the sink radius when 
\begin{equation}
 E_{\rm AGN} > \frac{3}{2} m_{\rm gas} k_{\rm B} T_{\rm min}
\end{equation}
where $m_{\rm gas}$ is the gas mass within the sink radius and $T_{\rm min}$ is the minimum feedback temperature. $T_{\rm min}$ should be chosen to be at least $10^7$ K, the temperature above which line cooling is not very efficient, so that the resulting feedback is independent of the value of $T_{\rm min}$. In our simulations we adopt the fiducial value $T_{\rm min}=10^7$ K. 

This model allows us to phenomenologically reproduce the two extreme regimes in which AGN feedback is thought to work \citep{Sijacki:2007p1032}: 
\begin{enumerate}
 \item "Quasar mode": in the case of cold dense gas accretion, more energy is required to reach the energy threshold. A large amount of energy is
 accumulated and released in a burst when the energy threshold is reached.
 \item "Radio mode": in the case of hot diffuse gas accretion, less energy is required to reach the threshold. Energy is injected in the gas in a
 quasi-continuous fashion.
\end{enumerate}

In the next section we will show how the presence or lack of SMBHs and their feedback can heavily influence the structure of massive galaxies in clusters.

\section{Results}
\label{sec:results}

In this section, we compare the properties of the BCGs in our two simulations to show the strong differences between the two models. We also compare our results with the two observational samples of early-type galaxies at $z\approx 0$ and $z\approx 1$ analysed by \cite{2008ApJ...688...48V}. {We stress that BCGs are particular kinds of early-type galaxies, so we make an additional comparison with the BCG sample analysed by \cite{2011arXiv1104.1239B} to show that our results are robust.} A final comparison is made with the cosmological hydrodynamical simulations performed by \cite{Naab:2009p5731} (a massive early-type galaxy) and \cite{Feldmann:2010p1516} (the central galaxy of a group).

\subsection{Identification of the BCGs}

We identified galaxies in our simulations with the AdaptaHOP algorithm \citep{2004MNRAS.352..376A}, using the version implemented and tested by  \cite{2009A&A...506..647T}. This version of the algorithm allows to use the Most Massive Substructure Method (MSM) to identify halos as well as their substructures, along with their centers, virial radii and masses. Galaxy centers were identified by running the same halo finder using the star particles. In each of the analysed snapshots, we defined the BCG as the object with the largest stellar mass; we verified that such a definition of the BCG also selects the galaxy closest to the cluster halo center.

In both simulations, the BCGs are surrounded by fairly smooth stellar halos extending to beyond 50 kpc. Similar stellar halos around BCGs have been found in numerical simulations by other authors \citep{Puchwein:2010p763}, and are similar to the intracluster light observed in galaxy clusters \citep{Lin:2003p1820, Gonzalez:2007p916, Giodini:2009p4283, 2010ApJ...720..569R}. There is no clear separation between the BCG and the intracluster stars, so just for accounting and comparison purposes, a criterion should be chosen to decide how to separate the central galaxy from the stellar halo.  We consider as part of the BCGs all the regions where the stellar density is larger than $2.5\times 10^6$ M$_{\odot}/$kpc$^3$. This choice is justified by the fact that lowering the density threshold by a factor of two does not influence the values of the measured stellar masses by more than 5\%. We also checked that the mass estimates we obtain using this criterion are robust by comparing them with estimates obtained using two additional methods: (i) we substituted the 3D density threshold with an equivalent surface density threshold; (ii) we use a criterion similar to that used by \cite{Naab:2009p5731}, where the BCG is identified as the set of star particles enclosed in a spherical region of fixed physical radius. The variations in stellar mass were always less than 5\%, therefore we consider the masses obtained with the stellar density criterion as our fiducial values. 

\begin{figure}
    \includegraphics[width=0.5\textwidth]{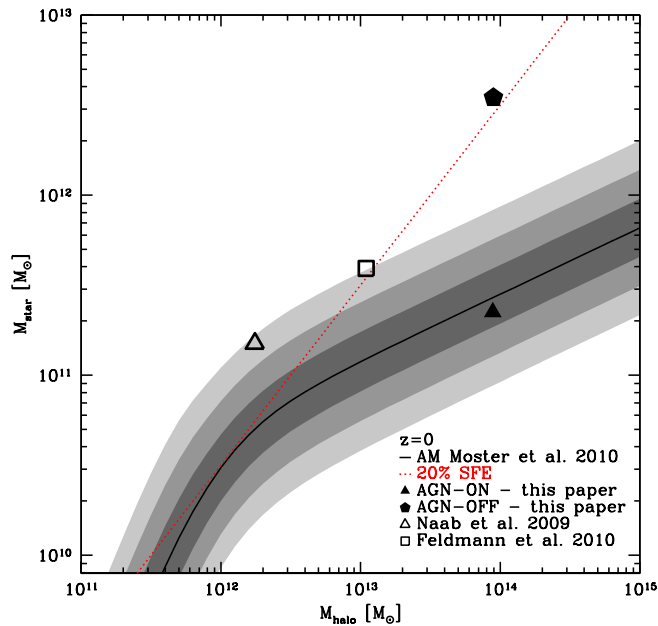}
  \caption{ Comparison of the stellar-vs-halo mass relation in 4 early-type galaxies from different cosmological simulations (filled and empty
  black dots). The red dotted line represents the relation expected for a 20\% star formation efficiency from the universal baryon fraction. The solid black
  line is the prediction from abundance matching \citep{Moster:2010p5423}. The grey shaded areas represent the $1\sigma$, $2\sigma$ and $3\sigma$
  scatter bars around the average relation. }
  \label{fig:AM}
\end{figure}

We find that the BCG mass at $z=0$ is ${\rm M_{\rm star}}=3.484\times10^{12}$ M$_{\odot}$ in the AGN-OFF run and ${\rm M_{\rm star}}=2.243\times10^{11}$ M$_{\odot}$ in the AGN-ON run. At $z=1$ we find ${\rm M_{\rm star}}=1.637\times10^{12}$ M$_{\odot}$ in the AGN-OFF run and ${\rm M_{\rm star}}=1.697\times10^{11}$ M$_{\odot}$ in the AGN-ON run. At both redshifts the total stellar mass of the BCG in the AGN-OFF simulations are almost 10 times larger. This fact implies that star formation quenching due to AGN feedback processes is very efficient in our model and has significantly changed the star formation history already by a redshift $z=1$.

In Figure \ref{fig:AM} we show the comparison of the stellar-vs-halo mass relation for the BCGs in our AGN-ON/OFF simulations with the predictions of 
abundance matching \citep{Moster:2010p5423} (black solid line). The BCG mass predicted by the AGN-OFF model deviates more than $3\sigma$ from the average relation obtained through abundance matching. Without feedback, the BCG forms stars at an efficiency of 20\% from the cosmic available baryon fraction, thus it lies close to the dotted red line. On the contrary, the AGN-ON model prediction is remarkably close to the average relation, and well inside the $1\sigma$ scatter bars. For comparison with recent simulations of massive early-type galaxies that do not include AGN feedback, we show the results of \cite{Naab:2009p5731} and \cite{Feldmann:2010p1516}. The mass of the central group galaxy simulated by these authors are both $\sim3\sigma$ above the average relation; also in this case the result is close to the simplified model with 20\% star formation efficiency. This suggests that AGN feedback could well be the mechanism that reduces the star formation efficiencies in massive galaxies. 

\begin{table*}\label{table:quantities}
{\bfseries BCG properties}
\begin{center}
\begin{tabular}{|c|c|c|c|c|c|c|c|}
\hline
\hline
 Simulation & Redshift & $M_{\rm star} [\hbox{M}_{\odot}]$ & $R_{\rm eff} [\hbox{kpc}]$ & $\sigma_{\rm eff}$ [km/s] & $\epsilon$ & $v/\sigma$ \\
\hline
\hline
AGN-ON & $z=0$ & $2.243\times10^{11}$ & 10.286 & 292.96 & $[0.123,0.155]$ & 0.08 \\
\hline
AGN-OFF & $z=0$ & $3.484\times10^{12}$ & 6.858 & 652.83 & $[0.257,0.498]$ & 0.65 \\
\hline
AGN-ON & $z=1$ & $1.697\times10^{11}$ & 8.254 & 277.93 & $[0.201,0.424]$ & 0.16 \\
\hline
AGN-OFF & $z=1$ & $1.637\times10^{12}$ & 4.784 & 525.32 & $[0.120,0.721]$ & 0.82 \\
\hline
\end{tabular}
\end{center}
\caption{ Properties of the BCGs. (1) Type of simulation. (2) Redshift $z$. (3) Stellar mass $M_{\rm star}$. (4) Half-mass radius $R_{\rm eff}$. (5) Velocity dispersion within $R_{\rm eff}$, $\sigma_{\rm eff}$. (6) Ellipticity $\epsilon$ at $R_{\rm eff}$. (7) Average $v/\sigma$ within $R_{\rm eff}$. }
\end{table*}

\subsection{Star formation rates}

\begin{figure}
    \includegraphics[width=0.5\textwidth]{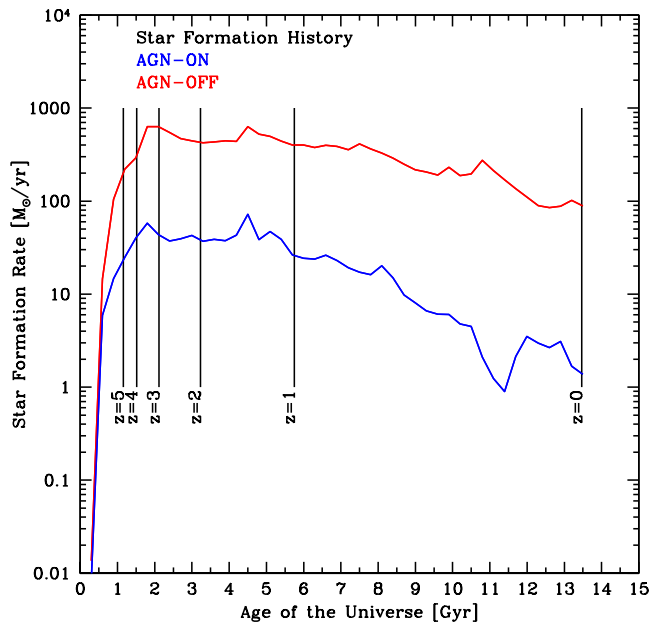}
  \caption{ Star formation rate in the BCG as a function of time. }
  \label{fig:time_hist}
\end{figure}

{The star formation history of the BCG galaxy in our simulations can be used to quantify the efficiency of the star formation quenching produced by our AGN feedback model. The difference between the star formation history between the AGN-ON and AGN-OFF simulations is striking. Fig. \ref{fig:time_hist} shows the rate at which stars found in the BCG at $z=0$ have been formed as a function of time. At the beginning of the simulation (age of the Universe $<1$ Gyr), when SMBHs have not formed yet the star formation rates in the AGN-ON (blue line) and AGN-OFF (red line) galaxies are almost indistinguishable. Later on AGN feedback starts to play its role in quenching star formation: the difference between the star formation rate in the two models is almost a factor 10, with the difference increasing towards $z=0$. \cite{2011arXiv1104.1626H} compare star formation rates of a wide variety of semi-analytical models and numerical hydrodynamical zoom simulations: they also find that the difference between the star formation rates in models including AGN and models do not including this effect is striking, however smaller than in our AGN-ON run. A similar result has been found by \cite{2011MNRAS.415.2782V}. A comparison with these results suggests that our AGN feedback model is particularly efficient in quenching star formation. It should be stressed that the such high feedback efficiency is expected because the resolution we reach is sufficient to resolve massive galaxies where SMBH reside already at $z\gtrsim6$, thus AGN feedback starts playing its role in quenching star formation at very early times. It is interesting to note how in both the simulations the star formation rate peaks before redshift $z=1$ and decreases toward $z=0$. The fact that star formation is extremely quenched between $z=1$ and $z=0$ in the AGN-ON model is also in agreement with the results of recent semi-analytical models of BCGs \citep{DeLucia:2007p1656}. Some starburst events are evident as peaks in the star formation rates at different times, while periods of strong AGN activity appear as sudden decreases of the star formation rate. This simple test shows how our simple AGN feedback model is indeed able to quench star formation, decreasing the stellar mass observed in the BCG.}

\subsection{Ellipticity and velocity distribution of stars in the BCGs} \label{subsec:ell_velstr}

\begin{figure*}
    \includegraphics[width=0.496\textwidth]{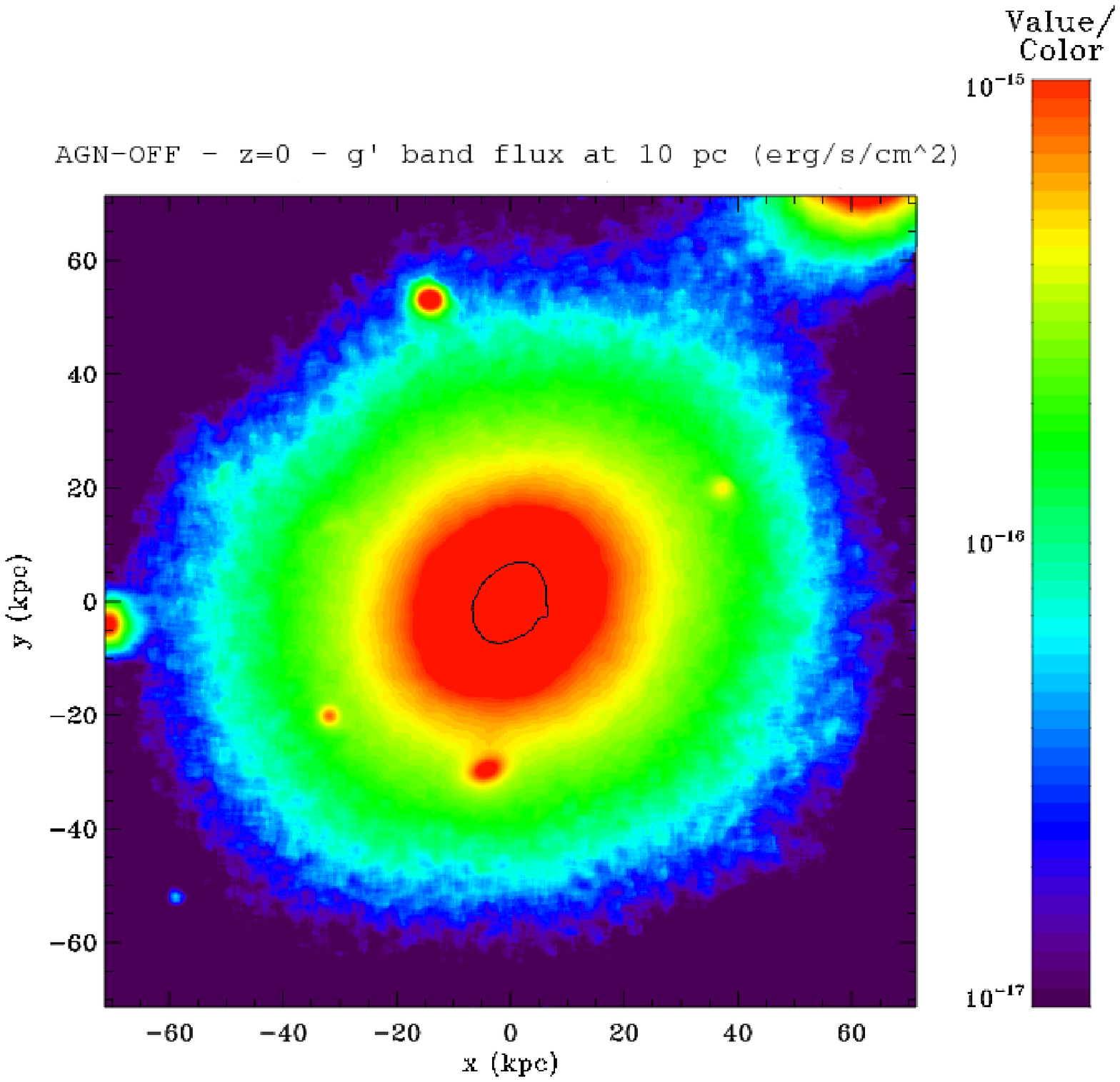}
     \includegraphics[width=0.496\textwidth]{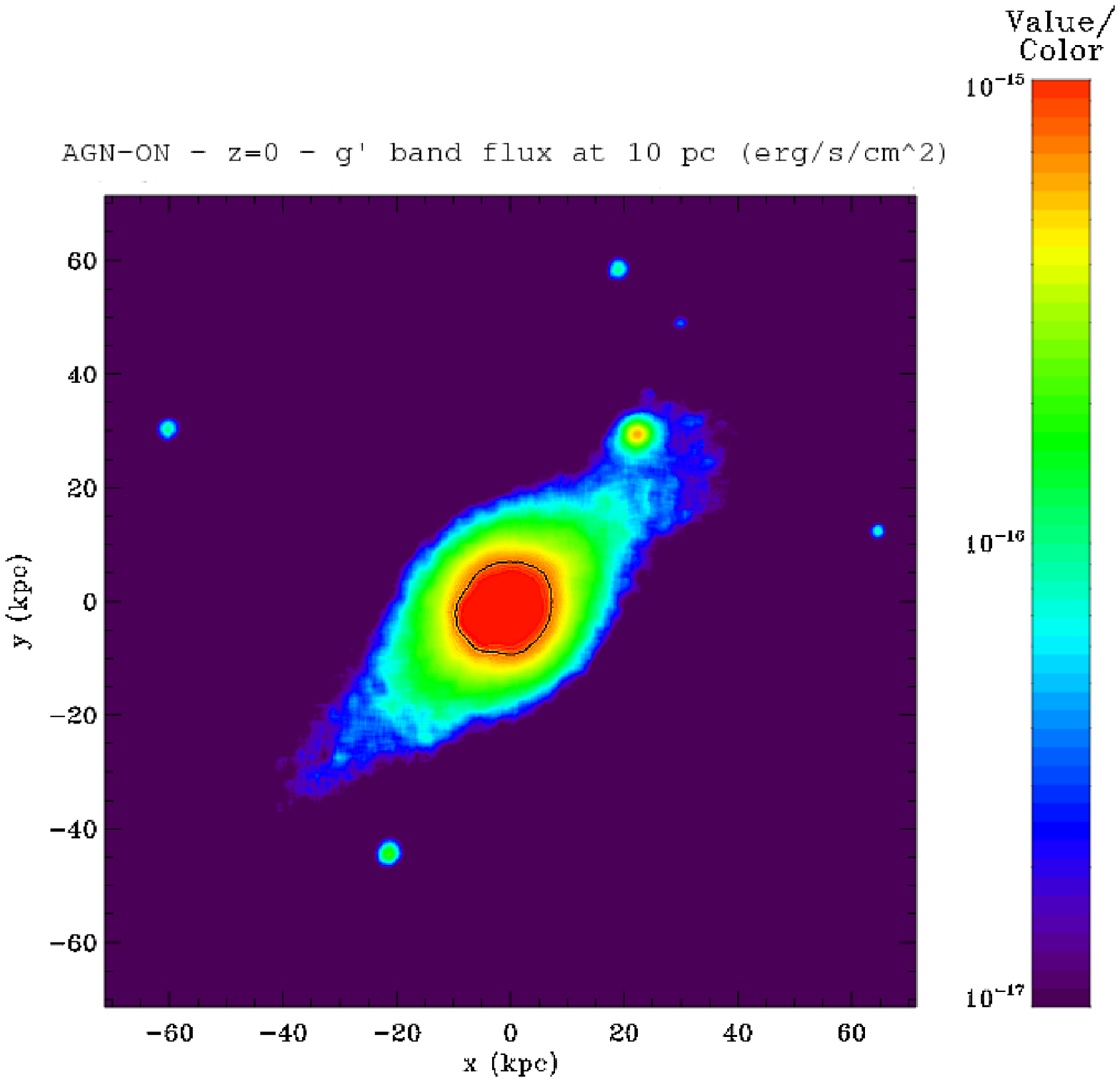}
    \includegraphics[width=0.496\textwidth]{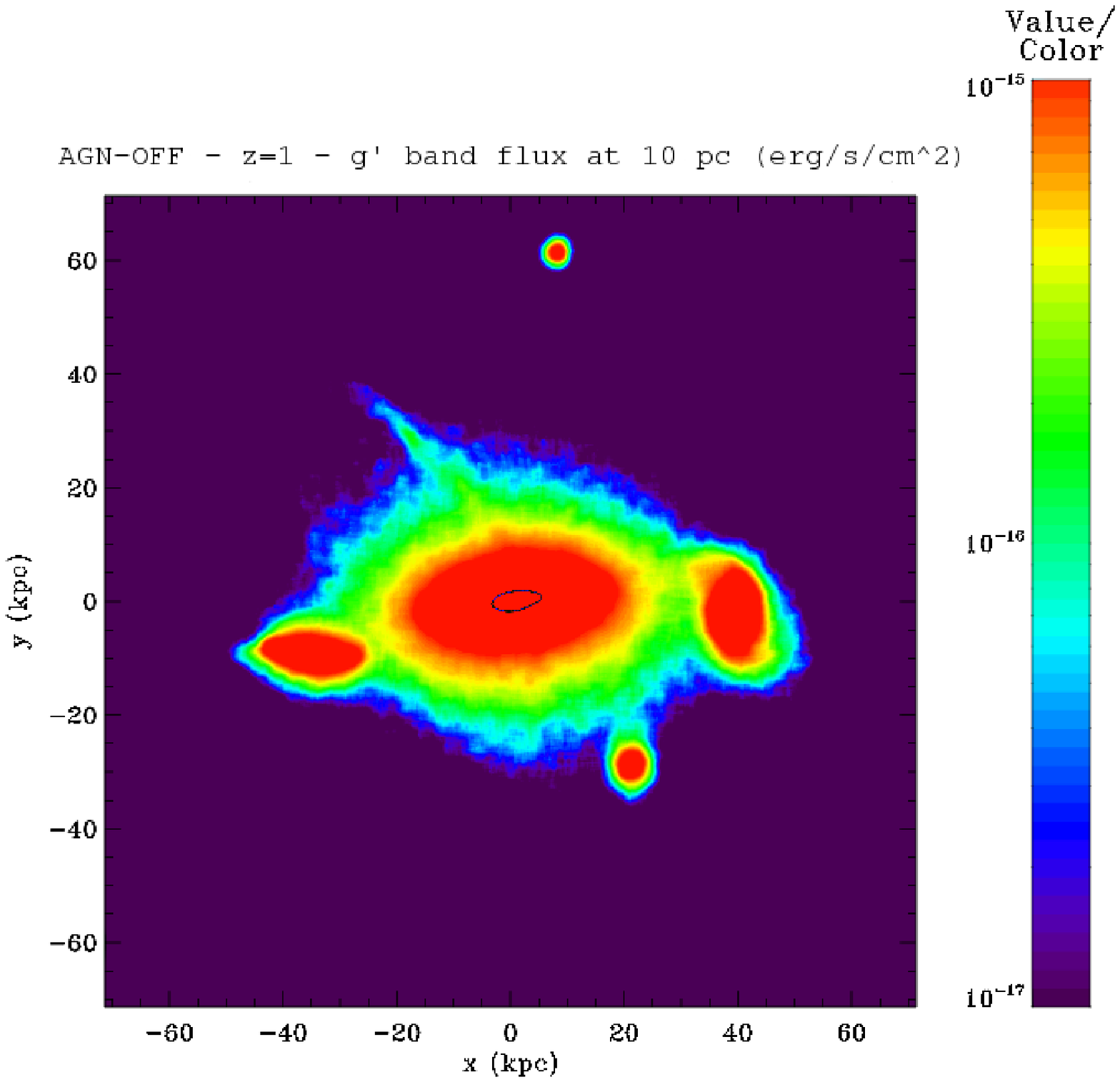}
     \includegraphics[width=0.496\textwidth]{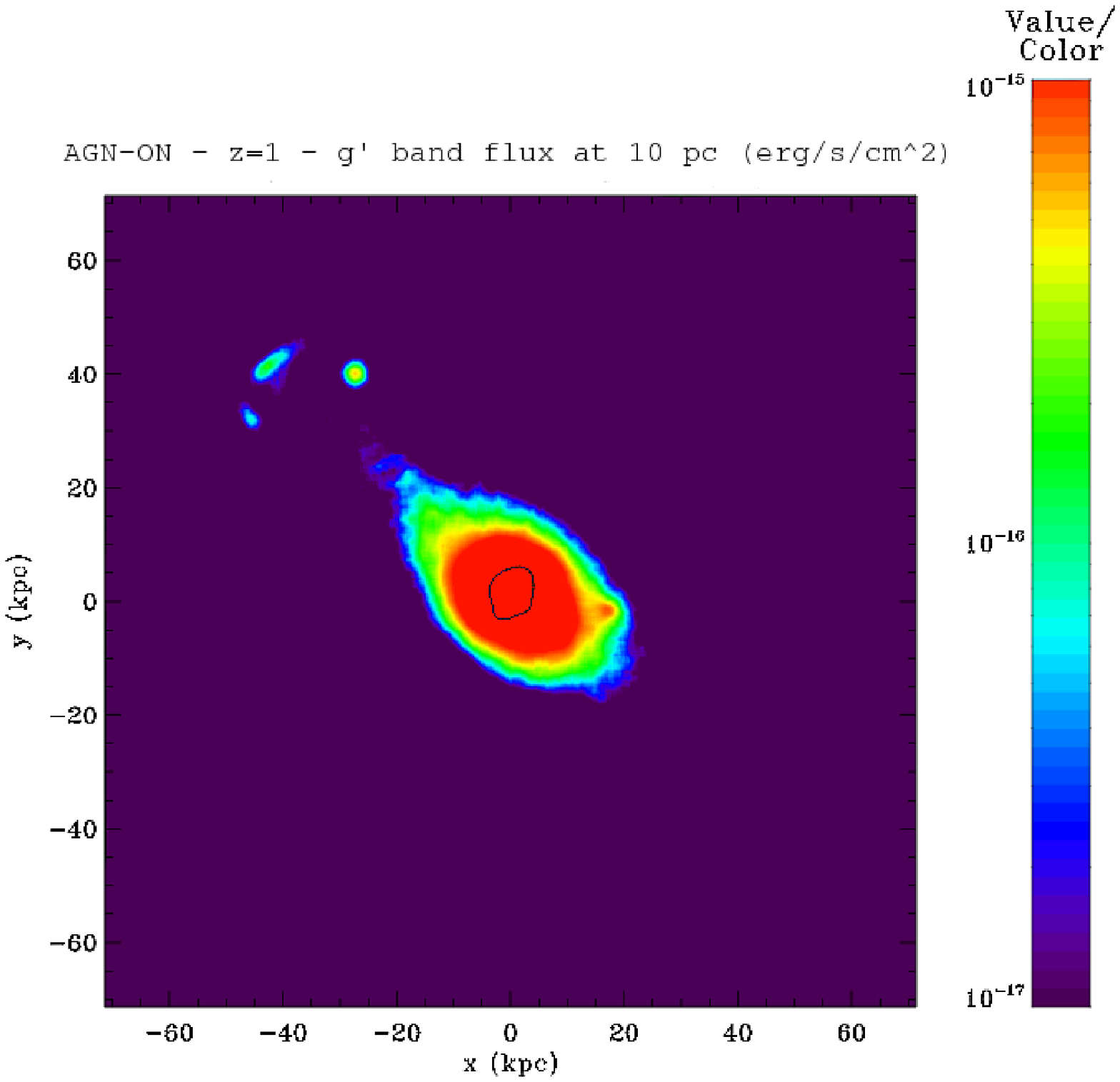}
 \caption{ Fluxes in the $g'$ band of the BCG projected along
 the $z$ axis of the periodic box. We filtered out all pixels with flux $<10^{-17}$ erg/s/cm$^2$. Top-left: $z=0$, AGN-OFF.
 Top-right: $z=0$, AGN-ON. Bottom-left: $z=1$, AGN-OFF. Bottom-right: $z=1$, AGN-ON. The black contours represent the half-light isophotes. }
  \label{fig:gbandmaps}
\end{figure*}

We constructed mock images of the central region of the simulated galaxy clusters to detect further differences between the two models. Figure \ref{fig:gbandmaps} shows a set of such  images at $z=1$ and $z=0$, where the color of each pixel represents the flux in the $g'$ band. We have chosen the same spatial and color scales in all the  pictures to show the remarkable difference in sizes between the AGN-ON and 
OFF simulations. The AGN-OFF BCG is much more extended than the AGN-ON one, at $z=1$ as well as at $z=0$, i.e. the global size of the galaxies scales with their mass. However, the half-light isophotes (black contours) show that the stellar light is more concentrated in the AGN-OFF BCG than in the AGN-ON one. This implies that the surface density profile of the AGN-ON galaxy is shallow when compared to the AGN-OFF one; this is indeed the case as we will show in \ref{subsec:profiles}.

To estimate the range of possible ellipticities an observer would see, we view the simulations from different angles and make the same surface brightness images. We then fit the half-light isophotes with ellipses and estimate their ellipticity $\epsilon=1-b/a$, where $b$ and $a$ are the semi-minor and semi-major axis of the ellipses. In this way we give the range of possible ellipticities for our galaxies. At $z=0$ the AGN-ON BCG has $\epsilon=[0.123,0.155]$, while the AGN-OFF BCG has $\epsilon=[0.257,0.498]$; both ranges are consistent with values typical of spheroidal systems, but the AGN-ON galaxy is slightly more spherical. At $z=1$ we have $\epsilon=[0.201,0.424]$ for the AGN-ON BCG, and $\epsilon=[0.120,0.721]$ for the AGN-OFF BCG; at this redshift, the AGN-ON galaxy is slightly more spheroidal than at $z=0$, while the ellipticity of the AGN-OFF galaxy is more typical of flattened ellipsoidal systems.

\begin{figure*}
    \includegraphics[width=0.496\textwidth]{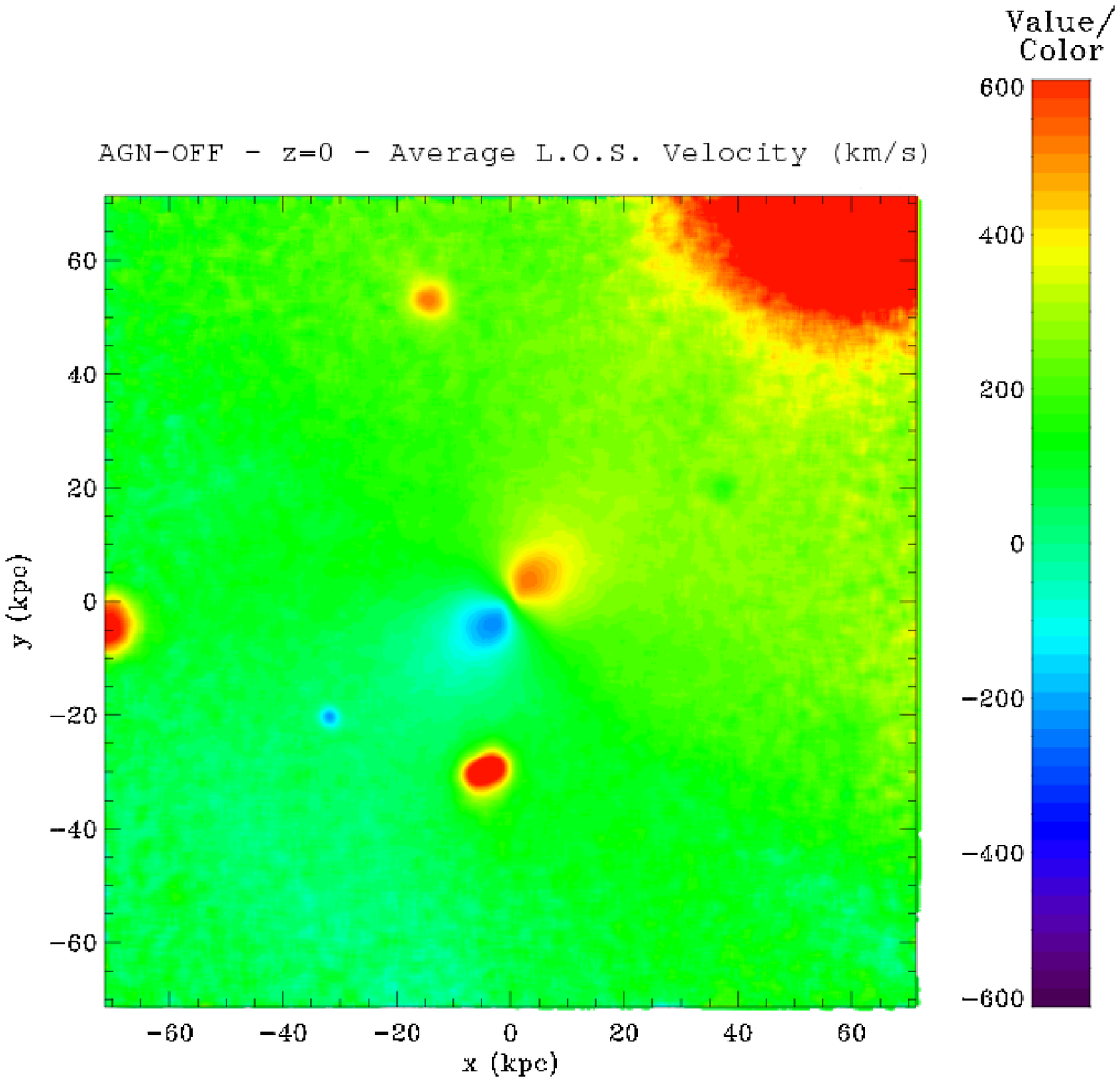}
     \includegraphics[width=0.496\textwidth]{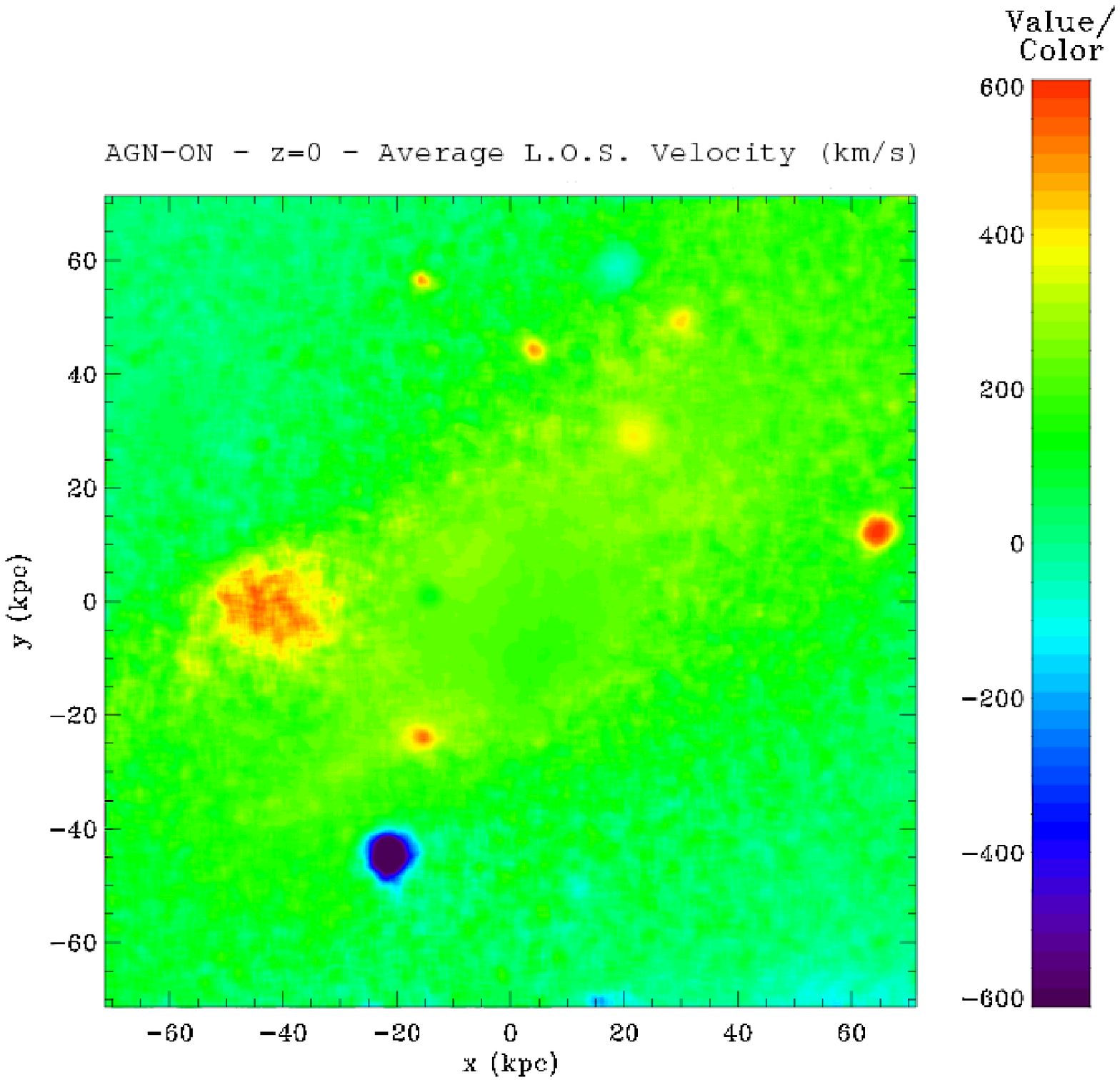}
    \includegraphics[width=0.496\textwidth]{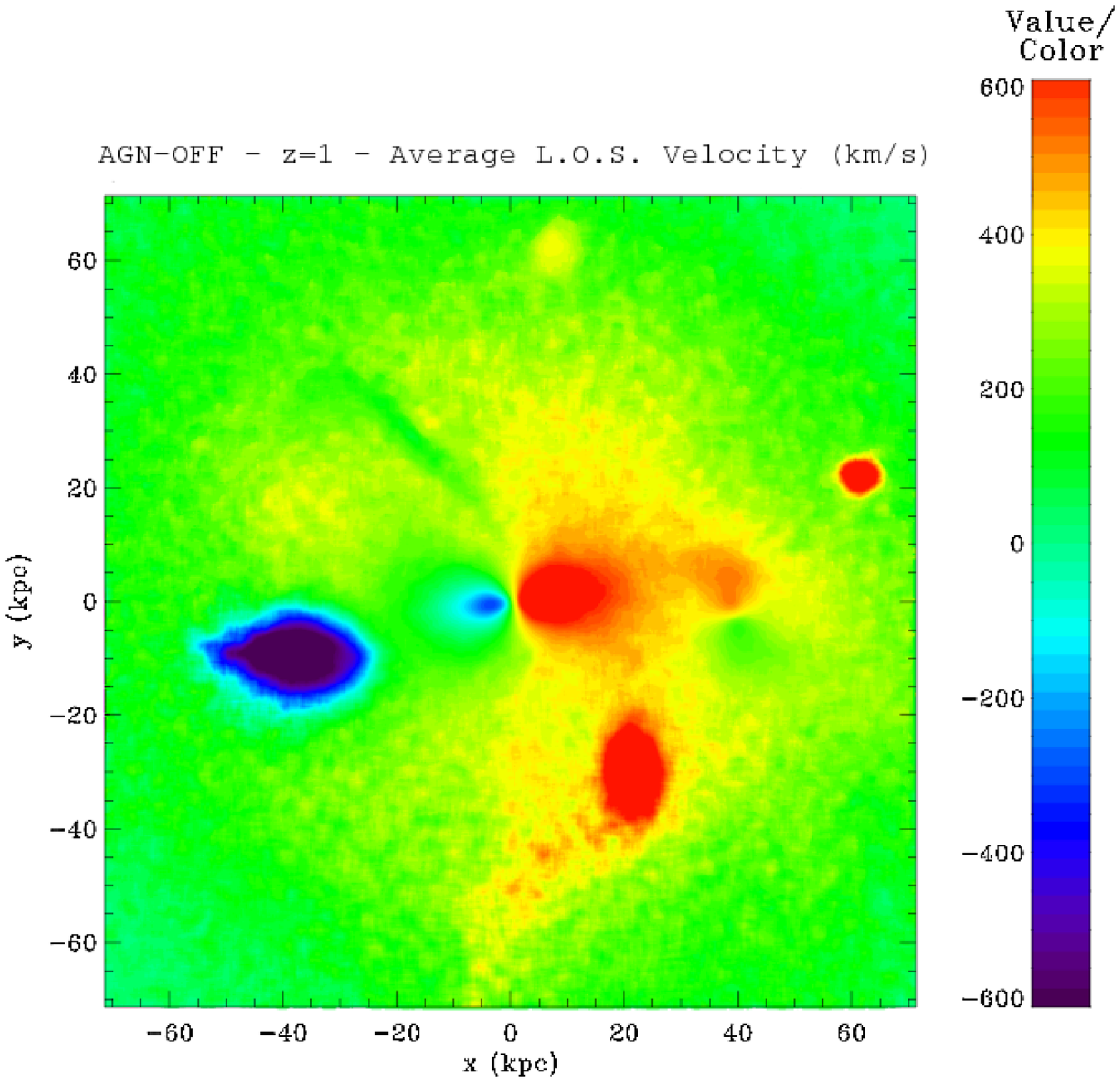}
     \includegraphics[width=0.496\textwidth]{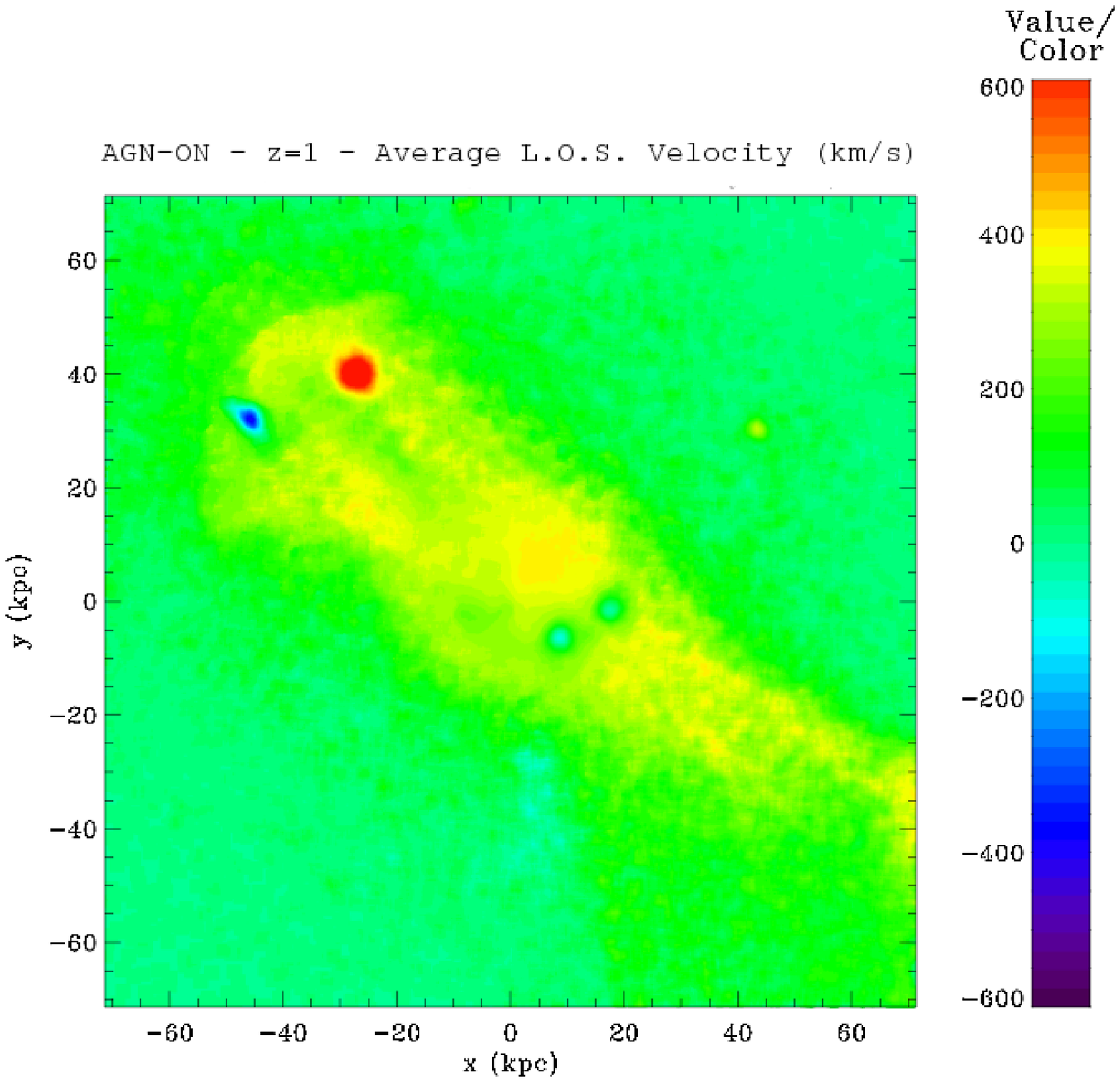}
 \caption{ Average velocity along the line of sight. The projection is along
 the $z$ axis of the periodic box. Top-left: $z=0$, AGN-OFF. Top-right:
 $z=0$, AGN-ON. Bottom-left: $z=1$, AGN-OFF. Bottom-right: $z=1$, AGN-ON. }
  \label{fig:velmaps}
\end{figure*}

\begin{figure*}
    \includegraphics[width=0.496\textwidth]{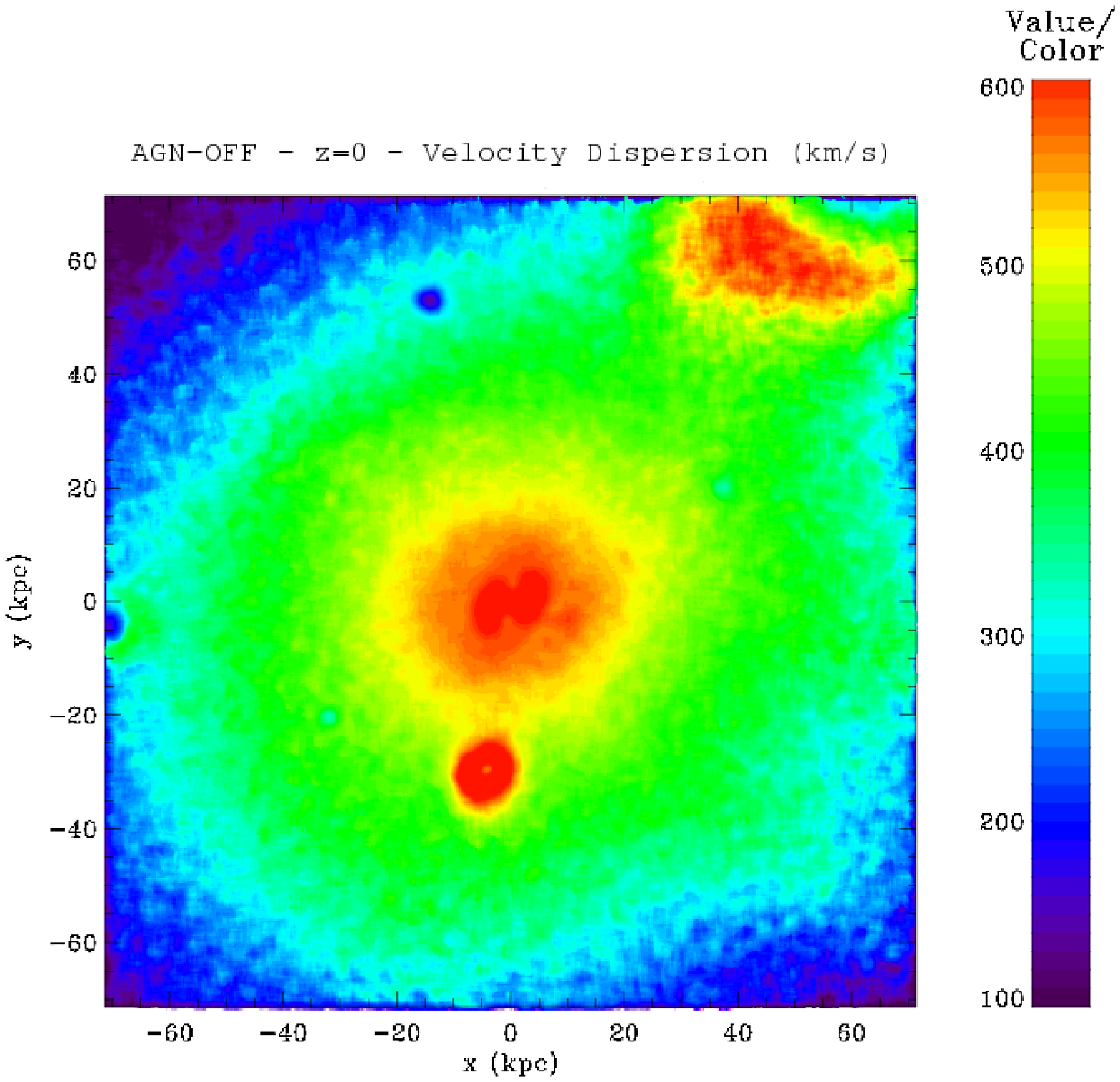}
     \includegraphics[width=0.496\textwidth]{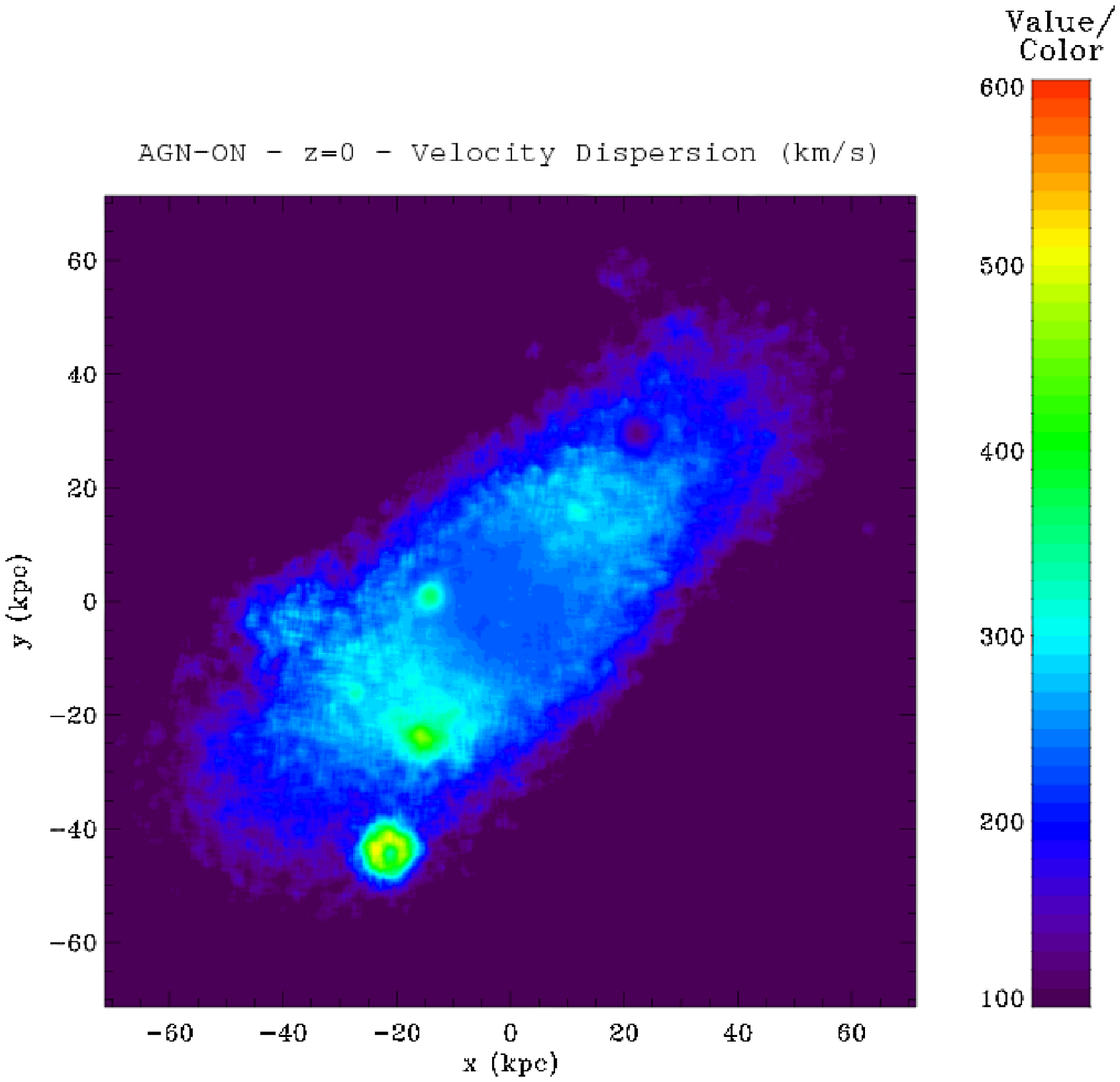}
    \includegraphics[width=0.496\textwidth]{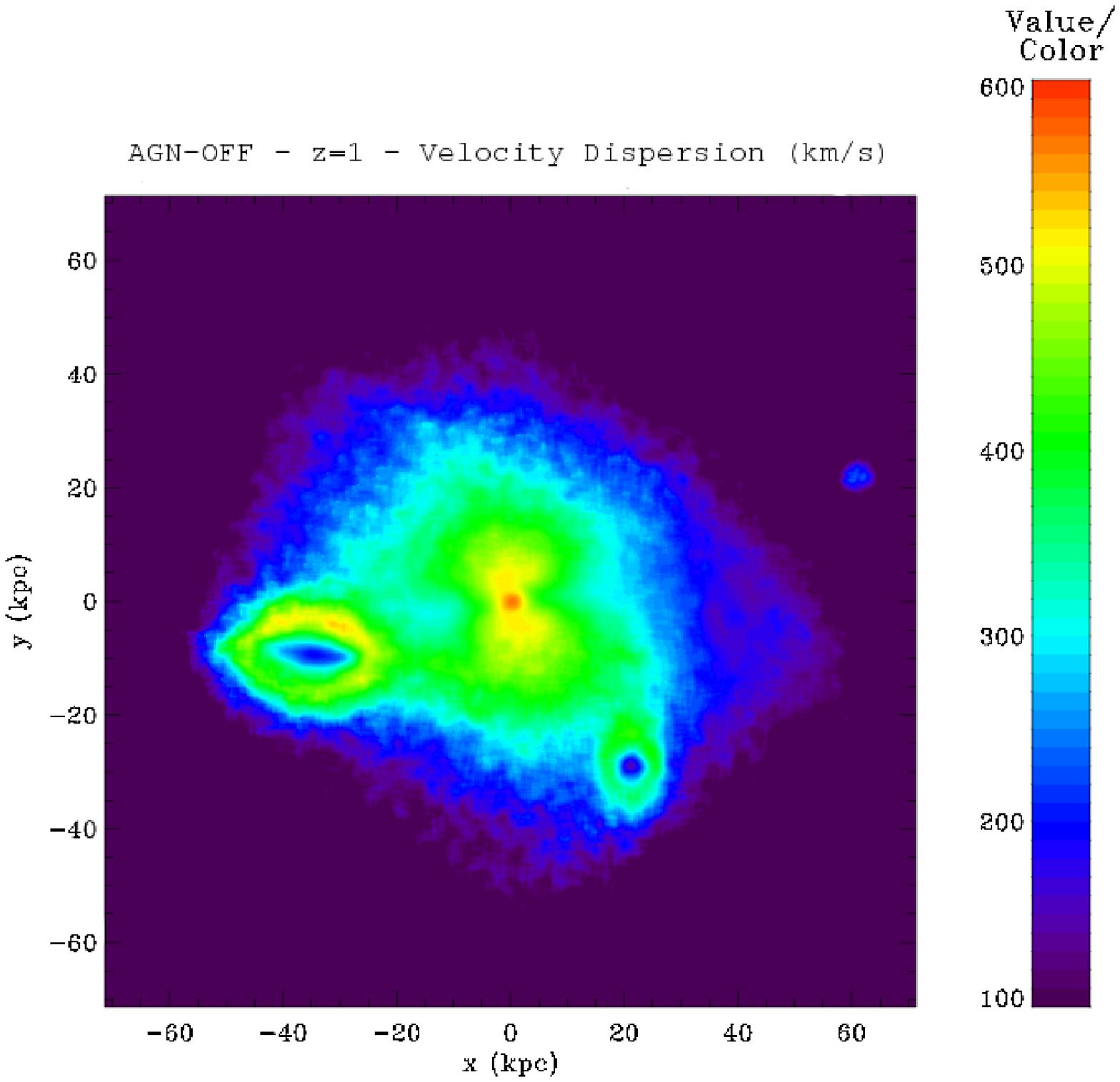}
     \includegraphics[width=0.496\textwidth]{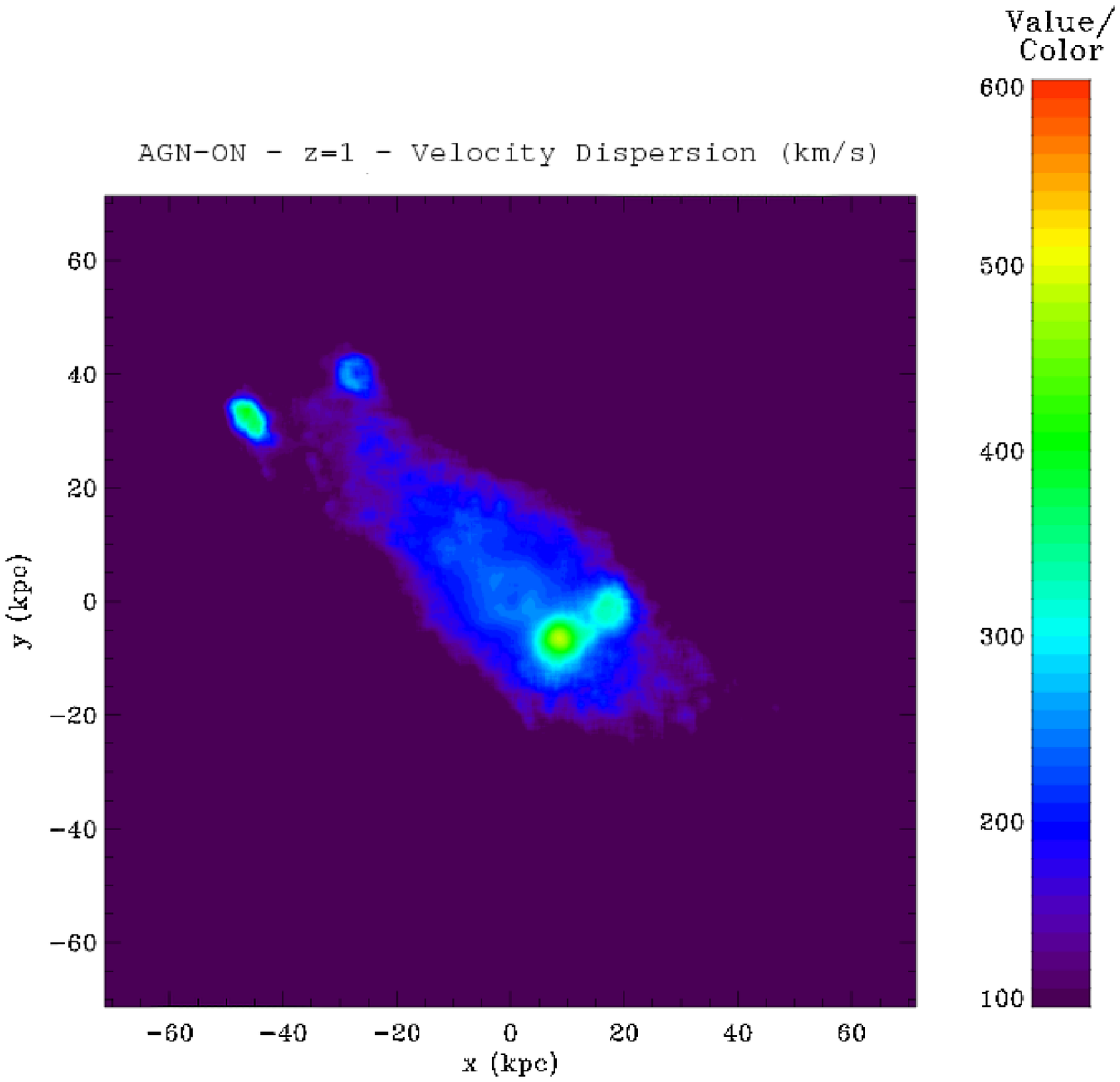}
 \caption{ Velocity dispersion along the line of sight. The projection is along
 the $z$ axis of the periodic box. Top-left: $z=0$, AGN-OFF. Top-right:
 $z=0$, AGN-ON. Bottom-left: $z=1$, AGN-OFF. Bottom-right: $z=1$, AGN-ON. }
  \label{fig:vdispmaps}
\end{figure*}

Figure \ref{fig:velmaps} shows the average line of sight velocity maps for the stars in the central region of the cluster at $z=0$ and $z=1$.  The AGN-OFF BCG hosts a rotating disk with a rotational velocity of a few hundreds km/s, both at $z=1$ and $z=0$. Interestingly, in the AGN-ON case there is no evidence for rotation at $z=0$, while at $z=1$ there is evidence for very slow rotation. These maps should be compared with those in Figure \ref{fig:vdispmaps} that show the line of sight velocity dispersion of the stars in the central region. The AGN-OFF velocity dispersion at $z=0$ is highly peaked in the center, where it is larger than $\sim 750$ km/s, and it decreases rapidly towards the external parts of the BCG. On the contrary, the AGN-ON velocity dispersion at $z=0$ does not peak in the center, but it increases towards the outskirts of the BCG (we will discuss this topic in greater detail in the next subsection). At $z=1$ the velocity dispersion is centrally peaked in both the simulations, but the values are at least a factor of two larger in the AGN-OFF case than in the AGN-ON case. If we rotate the galaxies edge-on and we account only for the regions within the half-light isophotes at $z=0$, we measure a mean $v/\sigma\sim 0.65$ for the AGN-OFF BCG, and $v/\sigma\sim 0.08$ for the AGN-ON BCG. At $z=1$ we find $v/\sigma\sim 0.82$ for the AGN-OFF BCG, and $v/\sigma\sim 0.16$ for the AGN-ON BCG.

It is interesting to compare the ($\epsilon,v/\sigma$) values with those observed in early-type galaxies, despite the fact that BCGs are a particular sub-category of these objects. \cite{ 2007MNRAS.379..418C} and \cite{2007MNRAS.379..401E} used 2D spectroscopy measurements from the SAURON survey \citep{2001MNRAS.326...23B, 2002MNRAS.329..513D} and analysed the orbital structure of 48 nearby S0 and E galaxies. They divide the galaxies in their sample into two groups: {\itshape slow rotators} with $\epsilon \lesssim 0.3$, $v/\sigma\lesssim 0.2$ and low specific angular momentum, and {\itshape fast rotators} with $\epsilon \lesssim 0.7$, $v/\sigma\gtrsim 0.2$ and a high specific angular momentum. Including the effects of AGN feedback in our simulations moved our BCG from the class of fast rotating early-type galaxies to that of the slow rotators.

The SAURON galaxies are not chosen to be central cluster galaxies, so we can also directly compare our simulations with the central galaxy in the Virgo Cluster, M87 (NGC 4486), that is also a slow rotator; according to \cite{2007MNRAS.379..418C}, M87 has $(\epsilon,v/\sigma)=(0.04,0.02)$, values that are consistent with the AGN-ON BCG simulation. This close match with M87 is interesting, given that in our simulations we only selected the Virgo-like halo based on its virial mass and merger history.

\subsection {Rotational velocity, velocity dispersion and stellar mass surface density profiles} \label{subsec:profiles}

\begin{figure*}
    \includegraphics[width=0.496\textwidth]{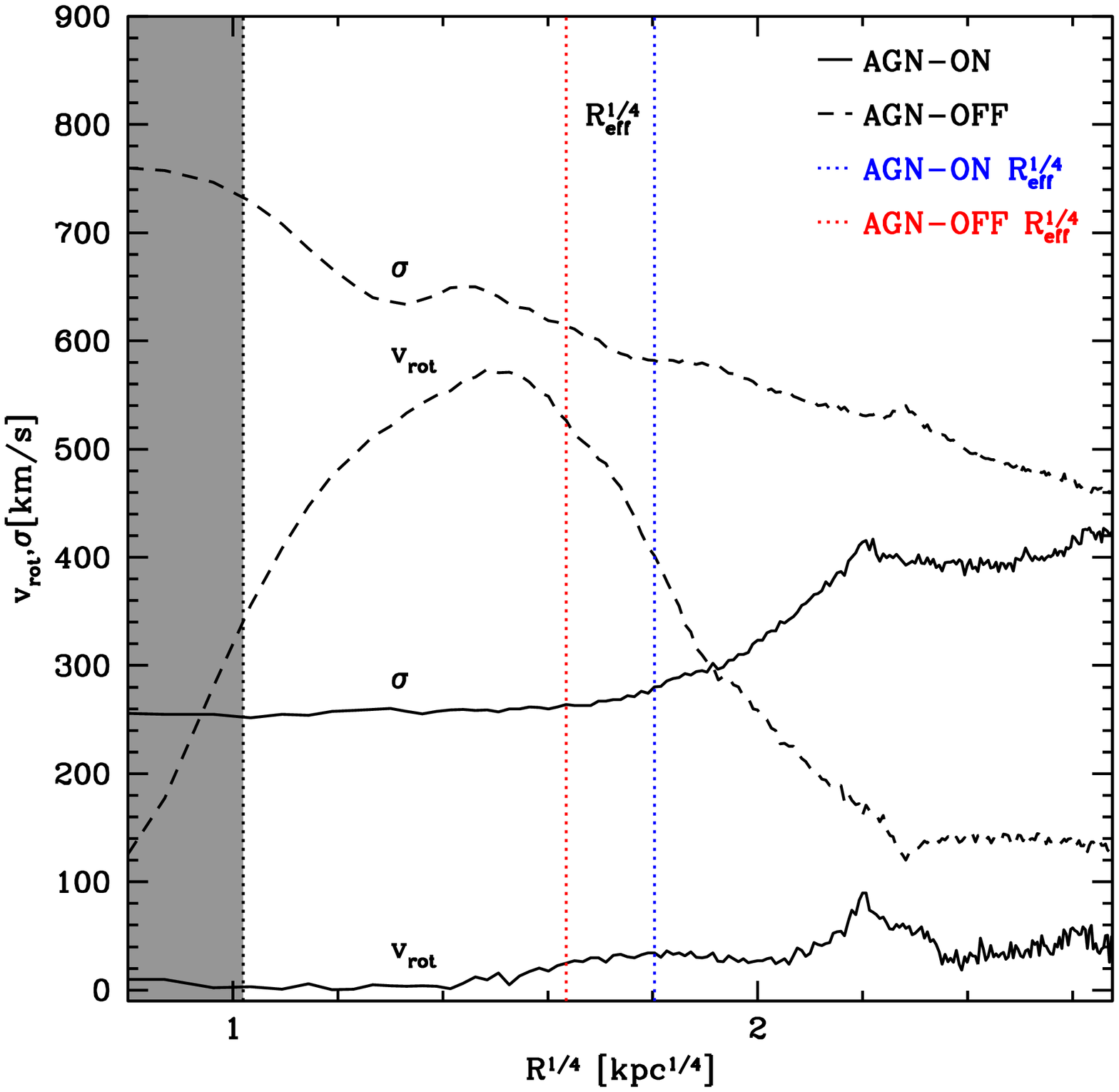}
    \includegraphics[width=0.496\textwidth]{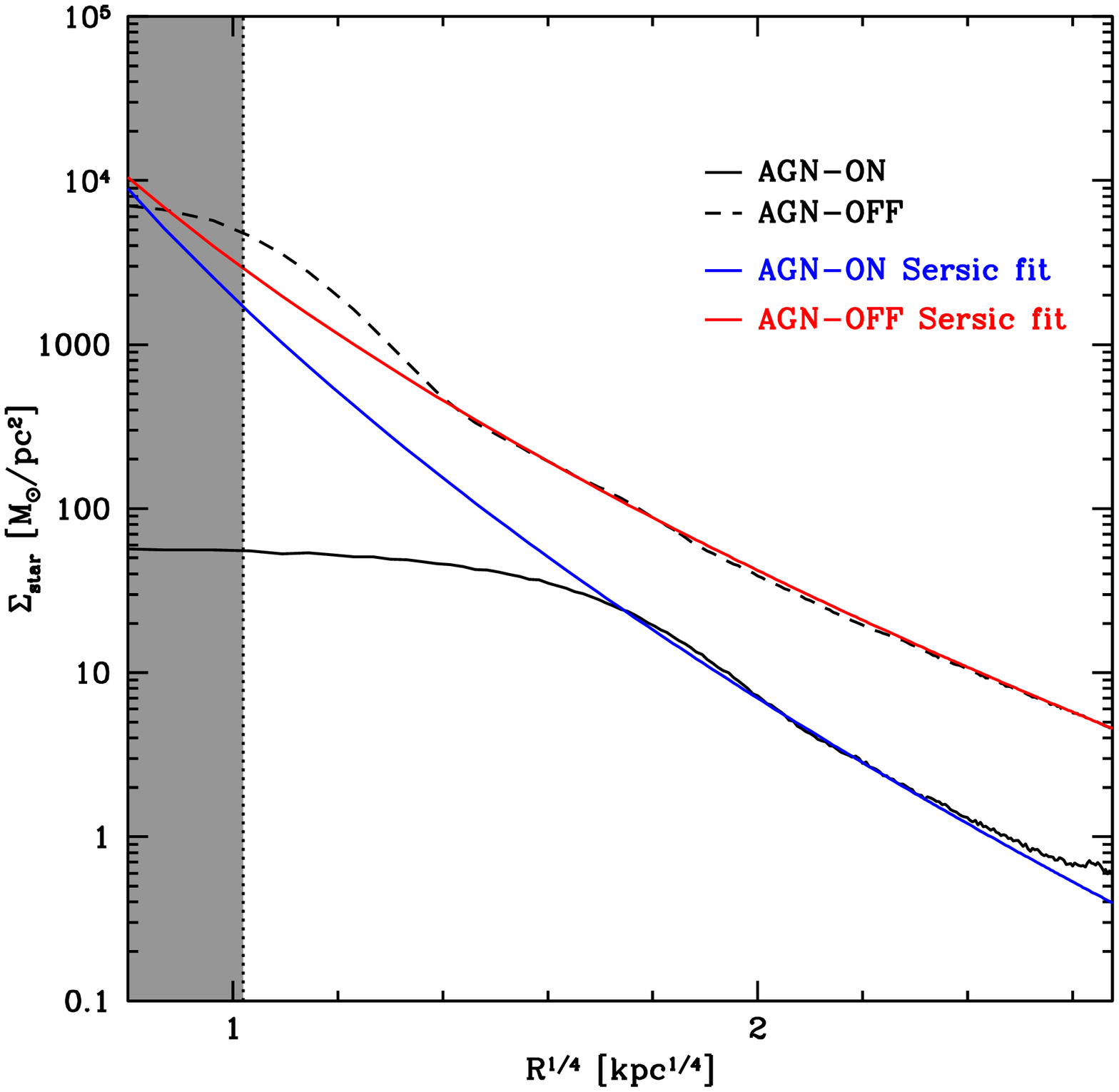}
  \caption{ Left panel: stellar velocity dispersion and rotational velocity of the BCG in our two models. The black solid lines are for the AGN-ON model and 
  the black dashed lines are for the AGN-OFF model. The dotted lines indicate the effective radii of the AGN-ON (blue) and AGN-OFF (red) BCGs.
  Right panel: stellar mass surface density profile of the BCG in our two models, compared with S{\'e}rsic profiles (see text for details). The black solid lines 
  are for tha AGN-ON model and the black dashed lines 
   are for the AGN-OFF model. In both 
  panels the grey shaded area shows the unresolved region of our simulations. }
  \label{fig:kinematicsandsigma}
    \includegraphics[width=0.496\textwidth]{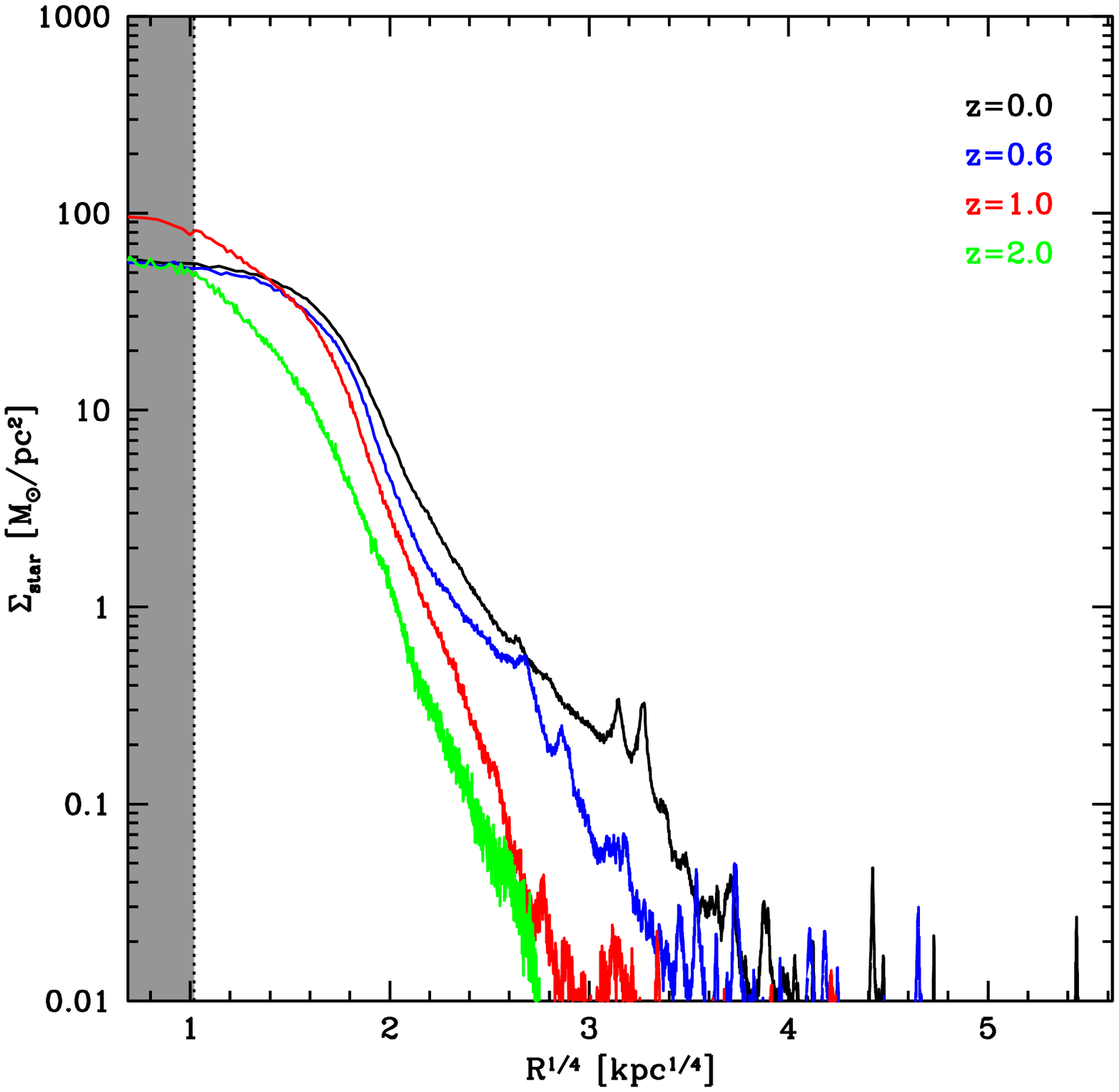}
    \includegraphics[width=0.496\textwidth]{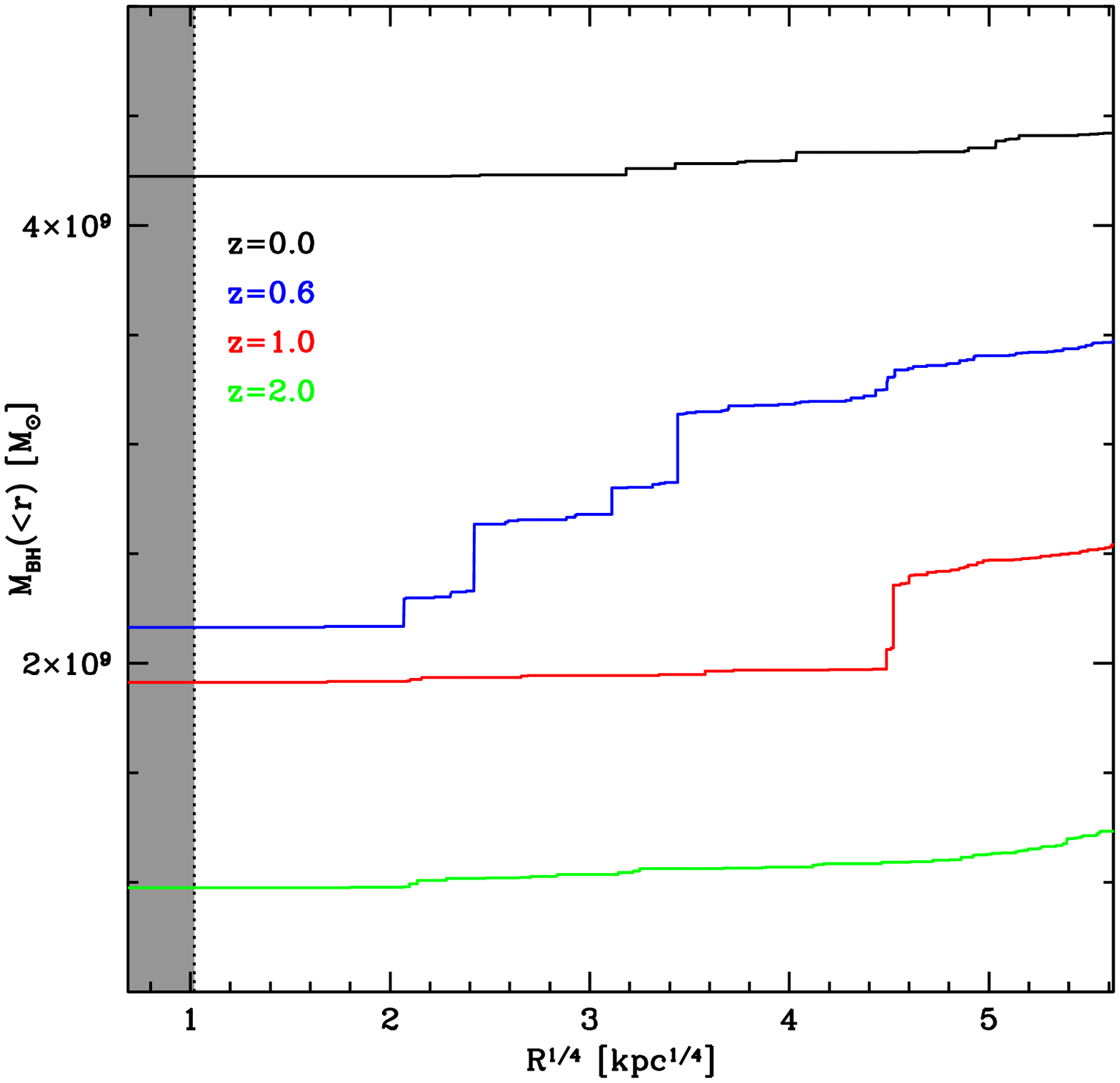}
  \caption{ Secular evolution of the stellar surface density profile towards a cored profile (left panel). The right panel shows the cumulative
  SMBH mass, showing that the late time evolution is governed by a succession of dry SMBH mergers towards the central galaxy, while, at early
  time, the total mass in SMBH is still growing by gas accretion.}
  \label{fig:scouringevol}
\end{figure*}

Here we analyse the circular velocity and velocity dispersion profiles of the two galaxies at $z=0$ (Figure \ref{fig:kinematicsandsigma}, left panel) to further explore the structural and kinematic properties of our BCG simulations. The rotational velocity (filled symbols) of the AGN-ON BCG within $R_{\rm eff}$ (blue dashed line) is never larger than 25 km/s and its value is slightly growing with the distance from the center. The line of sight velocity dispersion profile is nearly constant within the BCG, $250\lesssim\sigma\lesssim300$ km/s within $R_{\rm eff}$. Within this characteristic radius the galaxy is always dominated by random motions and has a very little rotation.

The AGN-OFF BCG has very different properties. Its velocity dispersion profile is maximum in the central region and decreases with radius; the rotational velocity increases with distance from the center, peaking just before $R_{\rm eff}$ is reached, then decreases again. Within $R_{\rm eff}$ the AGN-OFF galaxy is stil dominated by random motions, however, the rotational velocity is a significant fraction of the velocity dispersion ($\gtrsim 0.45$). Whilst rotating disk-like structures in fast rotators have been observed \citep{2008MNRAS.390...93K}, the peak $v_{\rm rot}$ value of our AGN-OFF simulation is a factor $\sim 2-3$ larger than the typical values measured for the fastest rotators. 

The velocity dispersion profiles measured for M87 \citep{2011ApJ...729..129M} are comparable with those of our AGN-ON BCG, at least beyond the spatial resolution limit. In SAURON the kinematic properties of M87 have been measured down to $\sim 0.1$ kpc \citep{2004MNRAS.352..721E}, but our spatial resolution limit is much larger: our AGN-ON model is not able to reproduce the velocity dispersion profile observed in M87 for $R<1$ kpc. However, when we consider the resolved region, $R>1$ kpc, the velocity dispersion profile of our AGN-ON model and that of M87 \citep{2004MNRAS.352..721E, 2011ApJ...729..129M} are rather similar.

On the right panel of Figure \ref{fig:kinematicsandsigma} we show the stellar mass surface density profiles at $z=0$ for our two BCGs. We find that these profiles can be fitted by S{\'e}rsic profiles for $R>10$ kpc for the AGN-ON BCG (blue line, S{\'e}rsic index $n=10$) and $R>4$ kpc for the AGN-OFF BCG (red line, S{\'e}rsic index $n=10$). We find that there is almost an order of magnitude difference between the stellar mass surface density of our two models. It is also interesting to compare the measured profiles with the S{\'e}rsic functions at $1 \hbox{ kpc}<R<10$ kpc: in the AGN-OFF case we find that there is a mass excess with respect to the S{\'e}rsic function and in the AGN-ON case we find a mass deficiency. On the observational side, similar results were described by several authors \citep{1999ASPC..182..124K, 2000ApJS..128...85Q, 2003AJ....125..478L, 2004AJ....127.1917T, 2005AJ....129.2138L, 2007ApJ...671.1456C,  2009ApJS..182..216K, 2011arXiv1108.0997G}. \cite{2009ApJS..182..216K} analyse the surface brightness profiles of elliptical and spheroidal galaxies in the Virgo cluster and show that it is possible to fit them with single light profiles to large radii. The fits are generally not satisfactory in the central regions; in particular, they find that most luminous elliptical galaxies have either light excesses or deficiencies at there centres. Light deficiencies are preferably found in the most luminous objects like NGC 4472 and M87 and the regions where they are found extend from the center to a distance $\lesssim 10$ kpc. If we assume that our surface density profiles can be mapped directly to surface brightness profiles, the observations of luminous galaxies are a much closer match to the AGN-ON simulation. In the next subsection we discuss the core formation process in more detail.

\subsection{Formation of the stellar core}

The stellar surface density profile of the AGN-ON BCG within the inner 10 kpc has constant density core. In the left panel of Figure \ref{fig:scouringevol} we show the time evolution of the stellar surface density profile of the AGN-ON BCG, from $z=2$ to  $z=0$. From this plot we can see that early on the stars are centrally concentrated, but a significant core starts developing at a redshift z=1. The core seems to be in place by a redshift z=0.6, or about 3 Gyrs, which is a cluster crossing timescale, but about 100 crossing times at the half light radius of 10 kpc. The core in the AGN-ON BCG doesn't form naturally, the AGN-OFF simulation has no such effect. Something has caused stars to move away from the inner region and this requires a significant amount of energy.

{\cite{Naab:2009p5731} showed that repeated minor dry mergers lead to large increases in the size of massive elliptical galaxies and to decreases of the central stellar density concentration. The cores of accreted galaxies act as perturbers and modify the stellar mass distribution through dynamical friction. We do not see this effect in our AGN-OFF run because completely dry mergers are rare due to the large reservoirs of gas residing in galaxies. In the AGN-ON case a significant fraction of the gas is removed from galaxies and the mechanism proposed by \cite{Naab:2009p5731} can be efficient. As we will show in Subsection \ref{subs:mass_size_vel}, we observe that the AGN-ON BCG is much more extended than the AGN-OFF BCG, in agreement with the \cite{Naab:2009p5731} picture, but it is challenging to produce an extended flat core like the one observed in our AGN-ON model only through repeated  dry minor mergers. SMBHs and AGN feedback provide additional processes able to contribute to the core formation mechanism.}

SMBH binaries are expected to form naturally in the hierarchy of mergers that lead to massive ellipticals and they provide several mechanisms that can produce cores. The most favoured model is {\itshape SMBH scouring} \citep{2003ApJ...596..860M}. At the galaxy centres the black holes form binary pairs that decay as they transfer energy to stars via three body encounters, thus they are able to eject stellar material from the central regions and form a core. Numerical experiments suggest that the cumulative effect of multiple SMBH dry mergers is able to remove a stellar mass 
that is $\sim 2-4$ times the final SMBH mass \citep{2007ApJ...671...53M}. 
This process is important from about 100 to 1 parsecs, so it is unresolved in our simulations which have 1 kpc force softening. At the softening length, the enclosed stellar mass is larger than the BH masses therefore a binary BH system can't form. However there is an addition process that we do resolve:
During the mergers the black holes would sink to the very central region due to dynamical friction. The numerical experiments presented in \cite{2010ApJ...725.1707G} show that the energy transferred from the sinking SMBHs to stars contributes to the formation of cored profiles; the typical mass deficits are found to have a similar magnitude as the SMBH mass.

The right panel of Figure \ref{fig:scouringevol} shows the cumulative SMBH mass from the center out to virial radius of the cluster at different output times. Each discontinuity in these profiles is associated to the presence of a SMBH. At early times the total mass in SMBHs is increasing mainly due to gas accretion, but at later times the growth is dominated by a succession of SMBH 
mergers. The mass of the central SMBH at $z=0$ is $M_{\rm BH}=4.2\times 10^9$ M$_{\odot}$. The simulations of \cite{2010ApJ...725.1707G} and \cite{2007ApJ...671...53M} used idealised  equilibrium models to study the effects of sinking massive binary objects. According to the model developed in \cite{2010ApJ...725.1707G}, a sinking SMBH of mass $M_{\rm BH}=4.2\times 10^9$ M$_{\odot}$ would produce a core of size $R_{\rm core}\sim3$ kpc. 

It should be stressed that these dynamical effects of sinking SMBHs act on all components of the mass distribution, although how stars, gas and dark matter respond can be quite different. Given this caveat, we note that the predictions for the mass deficit in such models is about one third lower than the mass deficit within the inner 10 kpc of our AGN-ON BCG, $M_{\rm def}=3.04\times 10^{10}$ M$_{\odot}$, measured as the difference between the mass predicted by the S{\'e}rsic profile in Fig. \ref{fig:kinematicsandsigma} and the actual mass enclosed in the same region. However, recent N-body experiments performed by \cite{2011arXiv1107.0517K} showed that when multiple (three or more) SMBHs are present, the core formation process is much more efficient: mass deficits in this case can be more than 5 times the total SMBH mass. These results bring the mass deficit obtained in pure N-body simulations with perturbing SMBHs much closer to the mass deficit observed in our AGN-ON run. 

Additional energy for the formation of the core can be provided by strong AGN driven outflows at $z<1$ that modify the local gravitational potential and may cause expansion of both the dark and stellar mass distribution. Similar processes have been observed in numerical cosmological simulations of dwarf galaxies in which gas outflows are generated by supernovae feedback \citep{1996MNRAS.283L..72N, 2010Natur.463..203G}. In the N-body simulations presented in \cite{1996MNRAS.283L..72N}, the mass outflows are simulated by growing and rapidly removing an idealised potential from the centre of an equilibrium realisation of a dark matter halo; the effect is the formation of a core in the density profile. In this model the efficiency of the process is $\propto M_{\rm disc}^{1/2}R_{\rm disc}^{-1/2}$, where $M_{\rm disc}$ is the mass of the disc and $R_{\rm disc}$ is its scale radius. This model is substantially different from ours, since we have multiple epochs of AGN driven gas outflows. However, these gravitational potential fluctuations should act in a similar way and could contribute to creating the large stellar cores we find.

\begin{figure*}
    \includegraphics[width=0.496\textwidth]{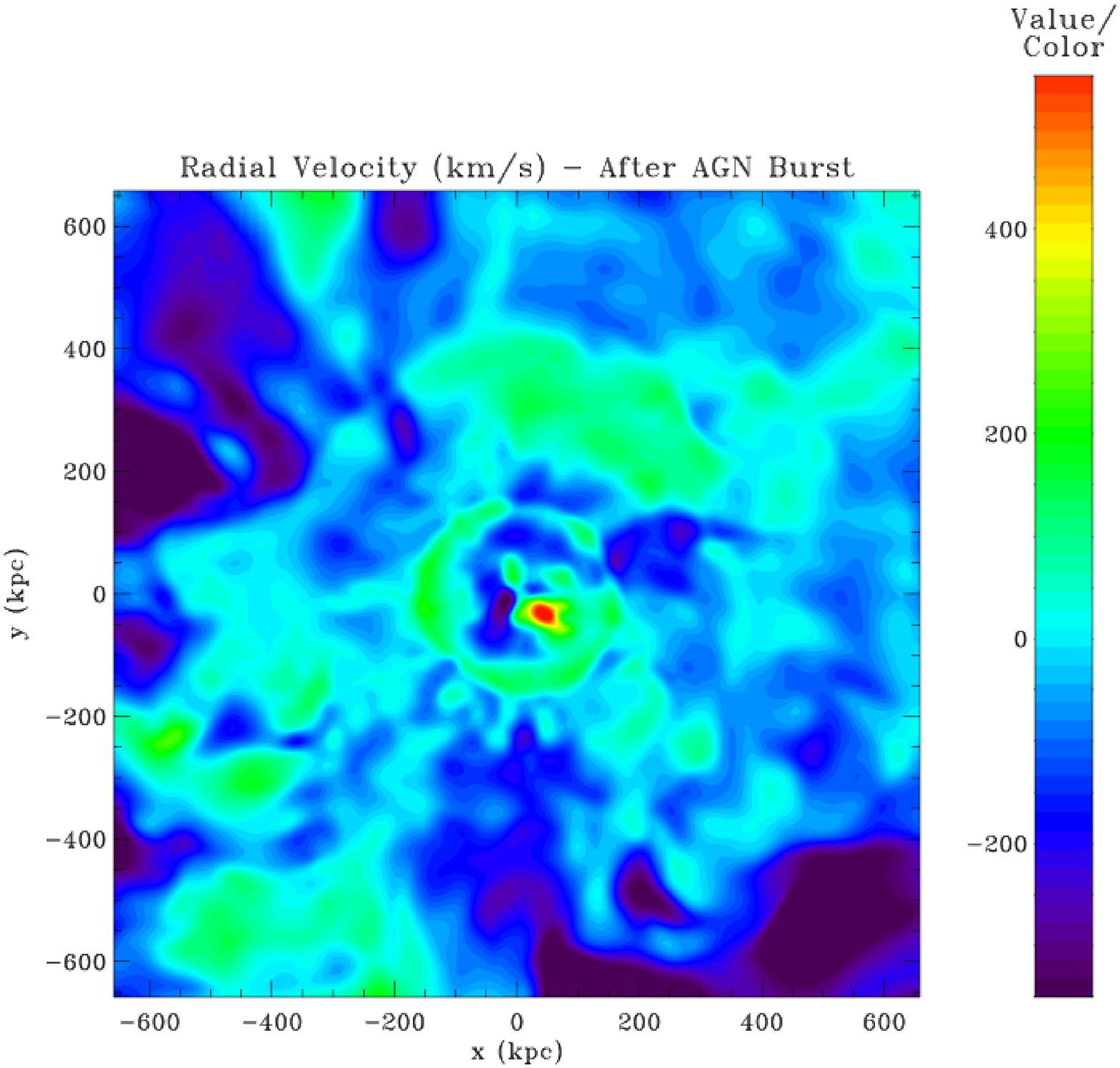}
     \includegraphics[width=0.496\textwidth]{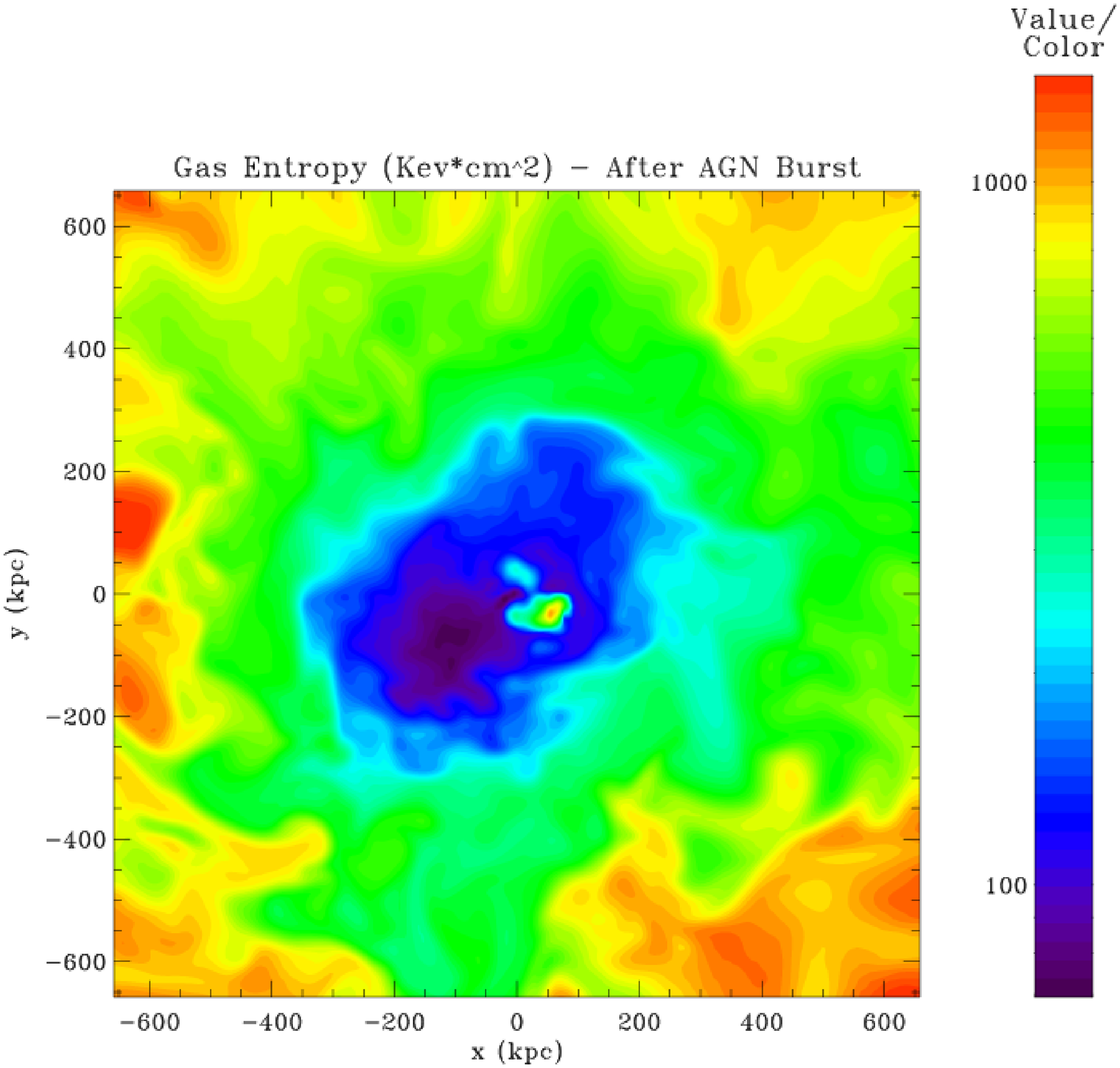}
 \caption{ Left panel: gas radial velocity map after an AGN burst in the AGN-ON simulation. Only the inner part of the cluster is shown. Positive values (green to red) of the radial velocity correspond to outflowing material, 
 negative values (blue to violet) are associated to inflowing material. Right panel: gas entropy map for the same region. }
  \label{fig:outflow}
\end{figure*}

{The maps in Fig. \ref{fig:outflow} show the structure of such outflows in our AGN-ON simulation. We show the gas properties just after a strong AGN burst at $z\sim 0.1$: the map in the left panel shows the radial velocity of gas with respect to the cluster centre (identified as the centre of the BCG), while the right panel shows a map of the gas entropy $K=k_{\rm b} T/n_{\rm e}^{2/3}$. These plots show  that in the outer regions of the cluster high entropy gas is fastly moving towards the center. This corresponds to hot intracluster medium cooling and flowing towards the cluster center. On the contrary, the inner regions ($r\lesssim$400 kpc) are characterized by convective motions pushing gas outwards at velocities $\sim 200$ km/s, e.g. note the ring-like structure observed in the radial velocity map. Looking at the entropy map we see that gas within 400 kpc from the centre is typically low entropy material. Observing the regions very close to the cluster centre and the BCG, we can speculate on the origin of the convective motions observed at higher distances. Close to the BCG a  significative fraction of the gas is falling towards the centre, but it is possible to observe an extended bubble of gas moving away from the centre at speed $\gtrsim 400$ km/s. This bubble has particularly high entropy and corresponds to gas that has been heated by a strong AGN burst. This gas will be eventually pushed away from the central region and will mix with the hot intracluster medium. The net effect of AGN activity is that gas is continuously expelled from the cluster inner regions.}

\begin{figure}
    \includegraphics[width=0.5\textwidth]{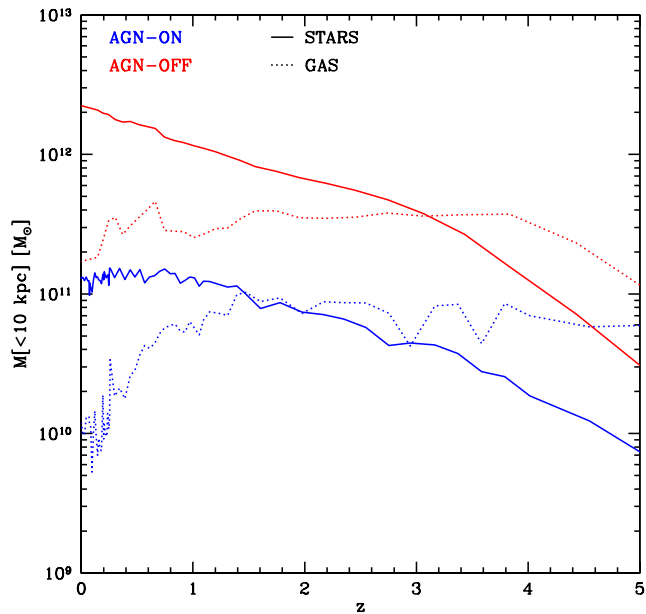}
  \caption{ Time evolution of the total stellar and gas mass enclosed within 10 physical kpc. }
  \label{fig:massevo}
\end{figure}

The plot in Figure \ref{fig:massevo} highlights the effect of gas outflows in our simulations: we plot the stellar (solid line) and gas (dashed line) mass enclosed within 10 physical kpc as a function of redshift for both the AGN-ON (blue) and AGN-OFF (red) simulations. There is more than an order of magnitude difference between the central stellar/gas mass in the two simulations. The plot shows that in the AGN-OFF case gas continuously accretes onto the central region, while the stellar mass is constantly increasing due to star formation events; the sudden increase of the gas mass are associated with galaxy mergers. A different scenario emerges in the AGN-ON simulation: the central gas mass is roughly constant at $z\gtrsim 1$, due to the fact that AGN feedback is efficient in heating and pushing gas away. In this regime star formation is still proceeding but it is strongly quenched. At $z\lesssim1$ AGN feedback in the central region becomes extremely efficient and $\sim 10^{11}$ M$_{\odot}$ of gas is expelled from the center before $z=0$. At $0\lesssim z\lesssim1$ the stellar mass inside the core is almost constant because there is very little star formation. The large gas outflows generate fluctuations in the gravitational potential that cause the ejection of stellar mass, possibly preventing the formation of a cusp in the surface density profile. 

The efficiency of core formation through such mass outflows has not been studied before. Naively applying the \cite{1996MNRAS.283L..72N} model under the assumption that $M_{\rm disc}$ equals the ejected gas mass and that $R_{\rm disc}=10$ kpc, it is possible to predict the formation of a core of size $R_{\rm core}\sim6$ kpc, which is close to the 10 kpc core observed in our simulations. 

Finally, we note that the gas mass within the cluster core of the AGN-ON simulation slowly decreases over time. This slow decrease in gas mass will lead to an adiabatic expansion of the total mass distribution, which will also contribute to the formation of a central core. Furthermore, the cooling time of hot gas within the inner 30kpc of the cluster centre is about one Gyr, thus we envisage that in its quiet mode the AGN can slowly eject the gas that rains down onto the centre from the inner cooling flow.

 To further test the efficiency of these processes, extensive numerical tests should be performed using {\itshape ad hoc} high resolution N-body simulations. We will go into further details about how these processes manifest themselves in galaxy clusters in a future paper (in preparation).

As mentioned in the previous section the size of the core in our AGN-ON BCG is slightly larger than what is expected for typical luminous elliptical galaxies. Also, the central surface brightness profiles of ellipticals in Virgo \citep{2009ApJS..182..216K} are seldom as flat as the stellar profile we observe in our AGN-ON BCG. This may suggest that the model we adopt to describe AGN feedback is slightly too efficient in injecting energy in the gas and in removing mass from regions surrounding the SMBHs.

\subsection{Masses, sizes and velocity dispersions }\label{subs:mass_size_vel}

We now compare the global properties of the galaxies with the observational data of \cite{2008ApJ...688...48V}. Their first sample is composed of $\gtrsim 17000$ early-type galaxies at redshifts $0.04<z<0.08$ taken from the SDSS database and whose stellar masses, effective radii (recovered from SDSS $g$-band imaging) and average velocity dispersion within
the effective radii are known. The second sample is composed of 57 cluster galaxies at redshifts  $0.8<z<1.2$, also with known stellar masses, effective radii (recovered from $z_{850}$-band imaging) and average velocity dispersions. Some of the most massive early-type galaxies in these samples are BCGs. To make our comparison with observations more focused on BCGs, we also compare our simulated central galaxies to the four BCGs at $z<0.09$ analysed by \cite{2011arXiv1104.1239B}.

For the two simulated BCGs we measured stellar masses, $M_{\rm star}$, half-mass radii, $R_{\rm eff}$\footnote{We assume that the radius that encloses half the mass of stars is close to the half light radii.}, and velocity dispersions within the half-mass radius, $\sigma_{\rm eff}$ at $z=0$ and $z=1$. The results are summarized in Table \ref{table:quantities}. The difference between the properties of our two BCGs is remarkable at $z=0$ as well as at $z=1$. We find that:
\begin{itemize}
 \item At $z=1$ the AGN-OFF stellar mass is an order of magnitude larger than in the AGN-ON case. The difference is even larger at $z=0$. \\
 \item The core of the stellar distribution, quantified by $R_{\rm eff}$, is almost twice as extended in the AGN-ON case than in the AGN-OFF case, both 
 at $z=1$ and $z=0$. \\
 \item The average velocity dispersion within $R_{\rm eff}$ is a factor $\sim 2$ larger in the AGN-OFF case at $z=0$ as well as at $z=1$. \\
\end{itemize}

The results of \cite{Naab:2009p5731} imply that the properties of very massive early-type are determined by repeated dry minor mergers that lead to the increase of galaxy sizes and masses between $z=1$ and $z=0$. In our AGN-ON model a large fraction of the gas is expelled from galaxies due to AGN feedback, increasing the probability of dry mergers. This effect is completely lacking in the AGN-OFF run because much more gas is present in all galaxies. This may partially explain why the AGN-ON BCG is more extended than its AGN-OFF counterpart. 

\begin{figure*}
    \includegraphics[width=0.496\textwidth]{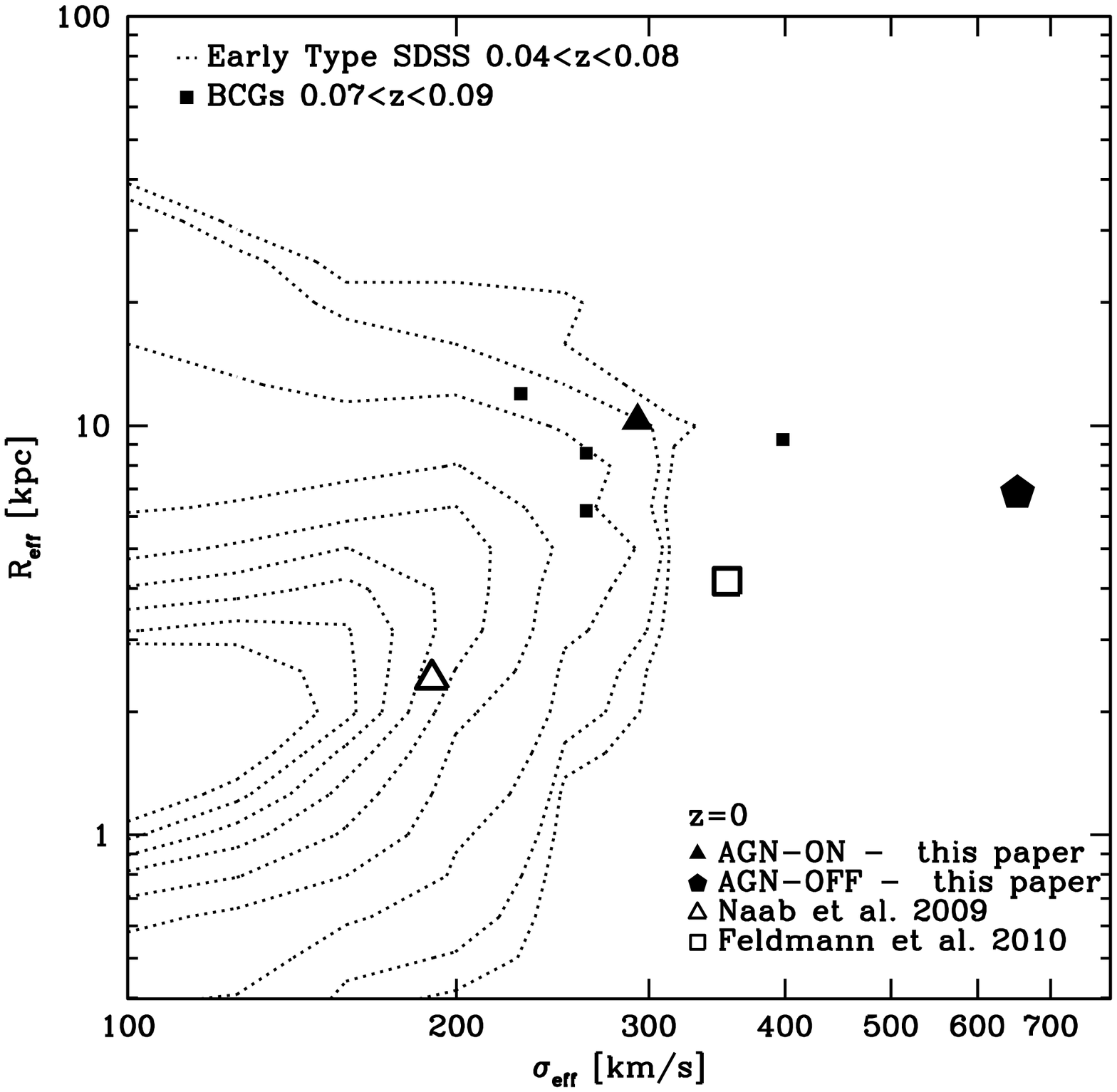}
     \includegraphics[width=0.496\textwidth]{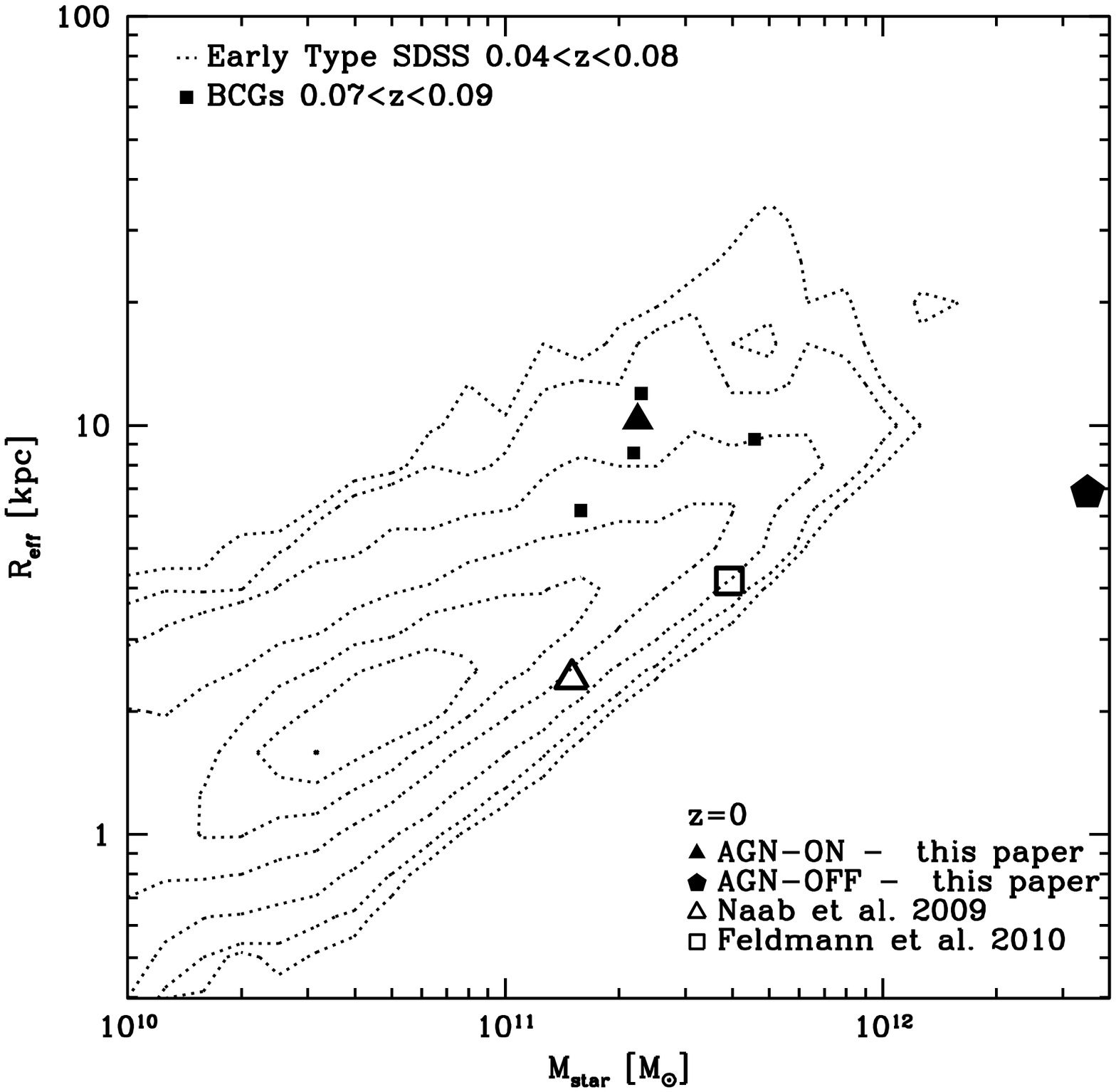}
 \caption{ Velocity-size (left panel) and mass-size (right panel) relation of early-type galaxies at redshift $z=0$ from Sloan data, compared to four
 early-type galaxies from different cosmological simulations. The black dotted lines are contours of the number of early-type galaxies per bin in the
 $0.04<z<0.08$ sample; going from outside-in we show contours for 5, 10, 30, 100, 200, 300, 400, 500, 600 galaxies per bin. Each bin has a size
 $\Delta \log(M_{\rm star})=\Delta \log(R_{\rm eff})=\Delta\log(\sigma_{\rm eff})=0.1$. The four BCGs analysed by Brough et al. 2011 are also shown as black 
 filled squares.}
  \label{fig:plane}
\end{figure*}

Figure \ref{fig:plane} shows the distribution of galaxies at low redshift in the ($\sigma_{\rm eff},R_{\rm eff}$) plane (left) and in the ($M_{\rm star},R_{\rm eff}$) plane (right). We compare the observations with our simulations and those of \cite{Naab:2009p5731} and \cite{Feldmann:2010p1516} (which do not include SMBHs nor AGN feedback). The simulation performed by \cite{Naab:2009p5731} reproduces quite well the properties of a typical early-type galaxy embedded in a Milky-Way sized halo at $z=0$. The central galaxy in the group simulated by \cite{Feldmann:2010p1516} has a stellar mass and effective radius compatible with several observed early-type galaxies, but a velocity dispersion slightly larger than expected. 

The AGN-OFF BCG of this paper is completely different from any observed low redshift massive early-type galaxy in the SDSS sample. Its mass and velocity dispersion are larger than those of any observed early-type galaxy by a factor $\sim3$ and $\sim2$, respectively. On the contrary, the properties of the AGN-ON BCG are quite consistent with those of observed galaxies. We point out that the properties of our AGN-ON BCG are also very similar to the four BCGs of \cite{2011arXiv1104.1239B}. These facts give strong observational support to galaxy formation models including AGN feedback. 
 
\begin{figure*}
    \includegraphics[width=0.496\textwidth]{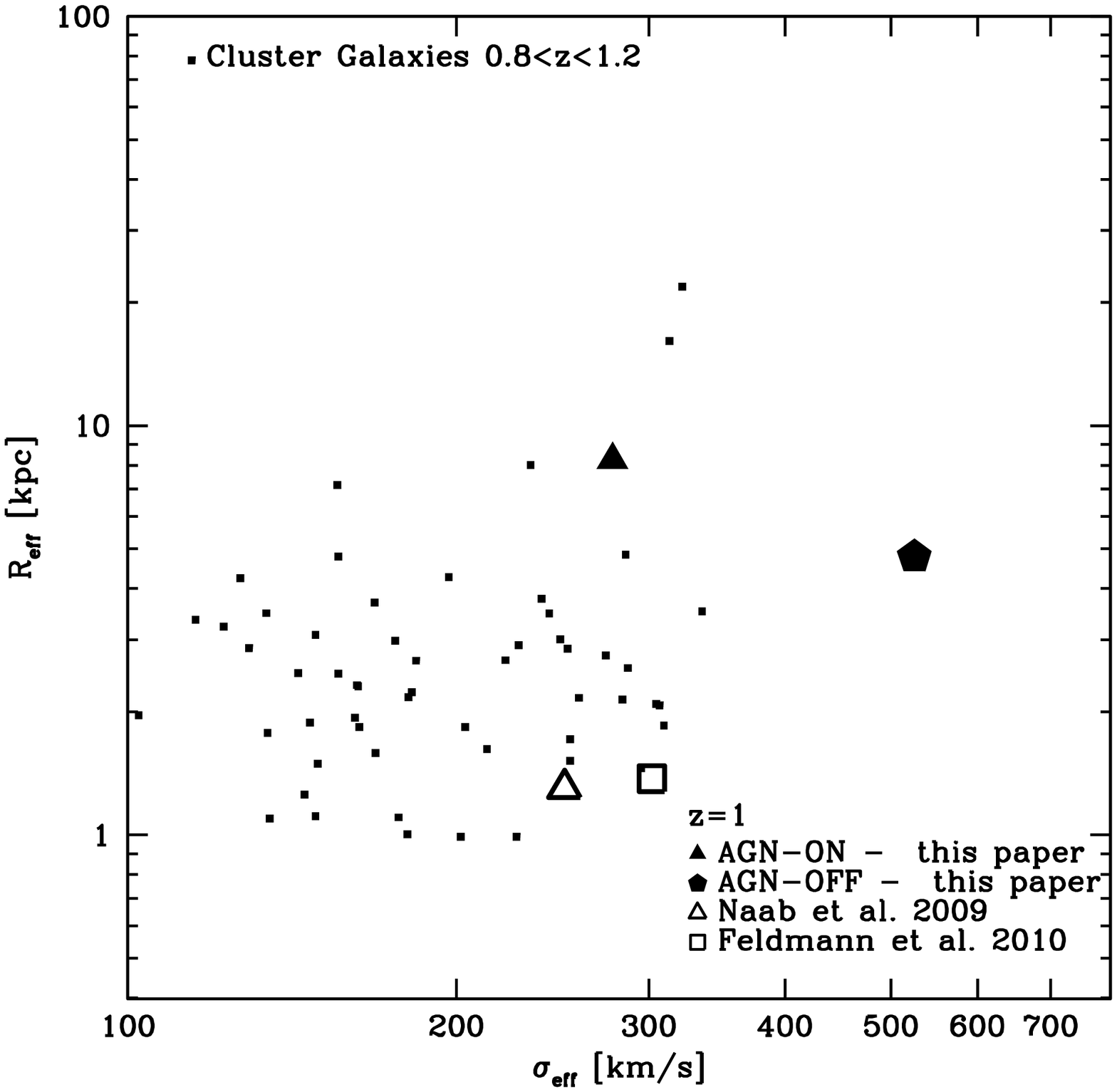}
     \includegraphics[width=0.496\textwidth]{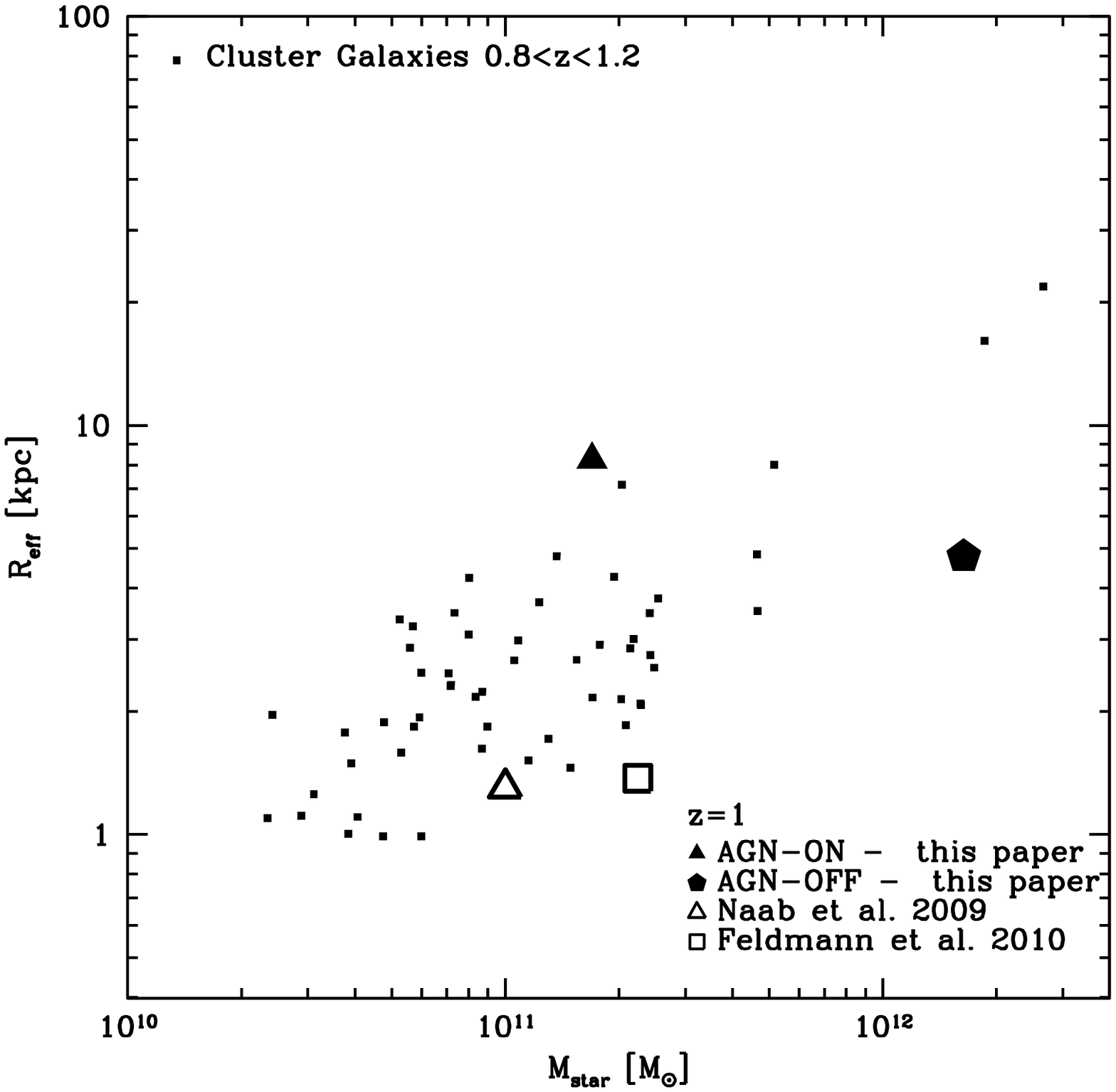}
 \caption{ Velocity-size (left panel) and mass-size (right panel) relation of clusters elliptical galaxies at redshift $z=1$ from van der Wel (2008),
 compared to 4 early-type galaxies from different cosmological simulations. }
  \label{fig:plane_z1}
\end{figure*}

To further test the effects of including the AGN we study the properties of the simulated galaxies at $z=1$. Figure \ref{fig:plane_z1} shows the distribution of the sample of cluster galaxies at $z\approx 1$ in the ($\sigma_{\rm eff},R_{\rm eff}$) plane (left) and in the ($M_{\rm star},R_{\rm eff}$) plane (right). For comparison, we also show also the results of \cite{Naab:2009p5731} and \cite{Feldmann:2010p1516}, even if they do not refer to galaxies in clusters. Again, we find that the AGN-OFF BCG does not match the observations. The properties of the AGN-ON BCG at $z=1$ are quite close to those of some of the most massive cluster galaxies in the sample, but the agreement is worse than at low redshift. 

\section{Summary and Conclusions}
\label{sec:summary}

We simulated the formation of a Virgo--sized galaxy cluster to study the properties of the massive bright galaxy that assembled at its centre. We considered two different models: the first one includes standard galaxy formation recipes (gas cooling, star formation and supernovae feedback) but no AGN feedback, whereas the second one also includes the effect of AGN feedback. Our results suggest that the effects of AGN feedback in clusters of galaxies is very important for the formation and evolution of the BCGs. {Whereas \cite{Naab:2009p5731} 
and \cite{Feldmann:2010p1516} show that AGN feedback is not strictly required to reproduce the properties of massive early-type galaxies in the field and in groups, we provide 
evidence that AGN feedback is needed to form realistic BCGs in clusters, i.e. in more massive halos.}

We compared the BCGs of our two simulations and we found substantial differences: when AGN feedback is neglected we obtain an elliptical galaxy whose properties (mass, size, kinematic structure, stellar density profile) are incompatible with observed elliptical galaxies at $z\sim 0 $ and $z\sim 1$. It is an extremely massive and fastly rotating galaxy, with a stellar cusp in the center. When AGN feedback is included the BCG appears to be completely different: it is 10 times less massive because star formation quenching is very efficient; it slowly rotates and its stellar surface density profile is cored in the inner 10 kpc. 

AGN feedback results in a stellar-to-halo mass ratio is consistent with the predicition of abundance matching \citep{Moster:2010p5423}. A comparison with the massive early-type galaxies in the samples analysed by \cite{2008ApJ...688...48V} shows that the mass, the velocity dispersion and the effective radius are consistent with those of the most massive early-type galaxies observed in the SDSS at $z\sim0$, and cluster galaxies at $z\sim 1$. We note that a slight decrease in the efficiency of AGN feedback would produce a slightly larger mass and a lower effective radius at $z=1$, bringing our simulated galaxy into an even closer agreement with the observations.

The existence of the core in the stellar surface density distribution is in agreement with what is observed for the most luminous and massive galaxies in the Virgo cluster that show significant mass deficiencies in their central regions \citep{2004AJ....127.1917T, 2007ApJ...671.1456C, 2009ApJS..182..216K, 2011arXiv1108.0997G}. We have discussed several mechanisms that could contribute to the shaping of the final properties of the BCG and, especially, to the formation of its core: (I) a series of dry mergers that lead to SMBHs sinking to the halo center via dynamical friction. This process can eject a large fraction of stars and dark matter from the central regions of the BCG \citep{2003ApJ...596..860M, 2010ApJ...725.1707G}. (II) AGN feedback driven gas outflows can  modify the gravitational potential in the regions close to SMBHs; these outflows are impulsive and the 'revirialisation' of the inner material can lead to the formation of a core \citep{1996MNRAS.283L..72N}. (III) The central hot gas slowly cools radiatively, falling onto the SMBH in a convective flow and is subsequently ejected impulsively. The slow loss of mass from the central region will result in the inner mass distribution expanding. The efficiency of each of these mechanisms will be explored using idealised numerical experiments in a subsequent study.

Observations show that low mass early-type galaxies typically have cusps in their surface brightness profiles, while high mass early-type galaxies preferably have centrally cored profiles \citep{2004AJ....127.1917T, 2007ApJ...671.1456C, 2009ApJS..182..216K}. We find that neglecting the presence of SMBHs and AGN feedback produces a cusp, while including these effects produces a core. These considerations suggests that there may be a close connection between the mass dichotomy in early-type galaxies and the presence of SMBHs. In high mass early-type galaxies the efficiency of the processes that lead to a core formation are expected to be higher than in lower mass early-type galaxies, thus lower mass galaxies may retain the cusps in the distribution of their stars.

\section*{Acknowledgments}
We thank our anonymous referee for helpful suggestions that greatly improved the quality of the paper. We also thank Lea Giordano for her suggestions about the topics discussed in this paper. We thank Robert Feldmann for providing us the group simulation data and Thorsten Naab for providing us the Milky-Way-sized simulation data. The AMR simulations presented here were performed on the Cray XT-5 cluster at CSCS, Manno, Switzerland.


\bibliography{papers}


\label{lastpage}
\end{document}